\documentclass[prx,aps,twocolumn,amsmath,amssymb,longbibliography]{revtex4-1}

\pdfoutput=1
\usepackage{hyperref}
\usepackage{graphicx}
\usepackage[usenames]{color}

\def\f#1{Fig.~\ref{#1}}

\def\s#1{Section~\ref{#1}}

\def\c#1{~\cite{#1}}

\def\cc#1{Ref.~\cite{#1}}

\def\beq{\begin{equation}}
\def\eeq{\end{equation}}
\def\bea{\begin{eqnarray}}
\def\eea{\end{eqnarray}}

\def\kt{k_{\rm B}T}

\begin{document}

\title{Learning to grow: control of material self-assembly using evolutionary reinforcement learning}
\author{Stephen Whitelam}\email{{\tt swhitelam@lbl.gov}}
\affiliation{Molecular Foundry, Lawrence Berkeley National Laboratory, 1 Cyclotron Road, Berkeley, CA 94720, USA}
\author{Isaac Tamblyn}\email{{\tt isaac.tamblyn@nrc.ca}}
\affiliation{National Research Council of Canada, Ottawa, ON, Canada}
\affiliation{Vector Institute for Artificial Intelligence, Toronto, ON, Canada}

\begin{abstract}

We show that neural networks trained by evolutionary reinforcement learning can enact efficient molecular self-assembly protocols. Presented with molecular simulation trajectories, networks learn to change temperature and chemical potential in order to promote the assembly of desired structures or choose between competing polymorphs. In the first case, networks reproduce in a qualitative sense the results of previously-known protocols, but faster and with higher fidelity; in the second case they identify strategies previously unknown, from which we can extract physical insight. Networks that take as input the elapsed time of the simulation or microscopic information from the system are both effective, the latter more so. The evolutionary scheme we have used is simple to implement and can be applied to a broad range of examples of experimental self-assembly, whether or not one can monitor the experiment as it proceeds. Our results have been achieved with no human input beyond the specification of which order parameter to promote, pointing the way to the design of synthesis protocols by artificial intelligence.
 
\end{abstract}

\maketitle

Molecular self-assembly is the spontaneous organization of molecules or nanoparticles into ordered structures\c{whitesides1991molecular,biancaniello2005colloidal,park2008dna,nykypanchuk2008dna,ke2012three,pfeifer2018synthetic}. It is a phenomenon that happens out of equilibrium, and so while we have empirical and theoretical understanding of certain self-assembling systems and certain processes that occur during assembly\c{doye2004inhibition,romano2011colloidal,glotzer2004self,doye2007condensed,rapaport2010modeling,reinhardt2014numerical, de2015crystallization,murugan2015undesired,whitelam2015statistical,nguyen2016design,whitelam2014self,jadrich2017probabilistic,lutsko2019crystals,fan2019orientational}, we lack a predictive theoretical framework for self-assembly. That is to say, given a set of molecules and ambient conditions, and an observation time, we cannot in general predict which structures and phases the molecules will form, and what will be the yield of the desired structure when (and if) it forms. As a result, industrial processes that use self-assembly, such as the crystallization of pharmaceuticals, require an empirical search of materials and protocols, often at considerable time and cost\c{chen2011pharmaceutical,threlfall2000crystallisation,rodriguez1999significance,shekunov2000crystallization}.
\begin{figure*}[] 
   \centering
\includegraphics[width=0.9\linewidth]{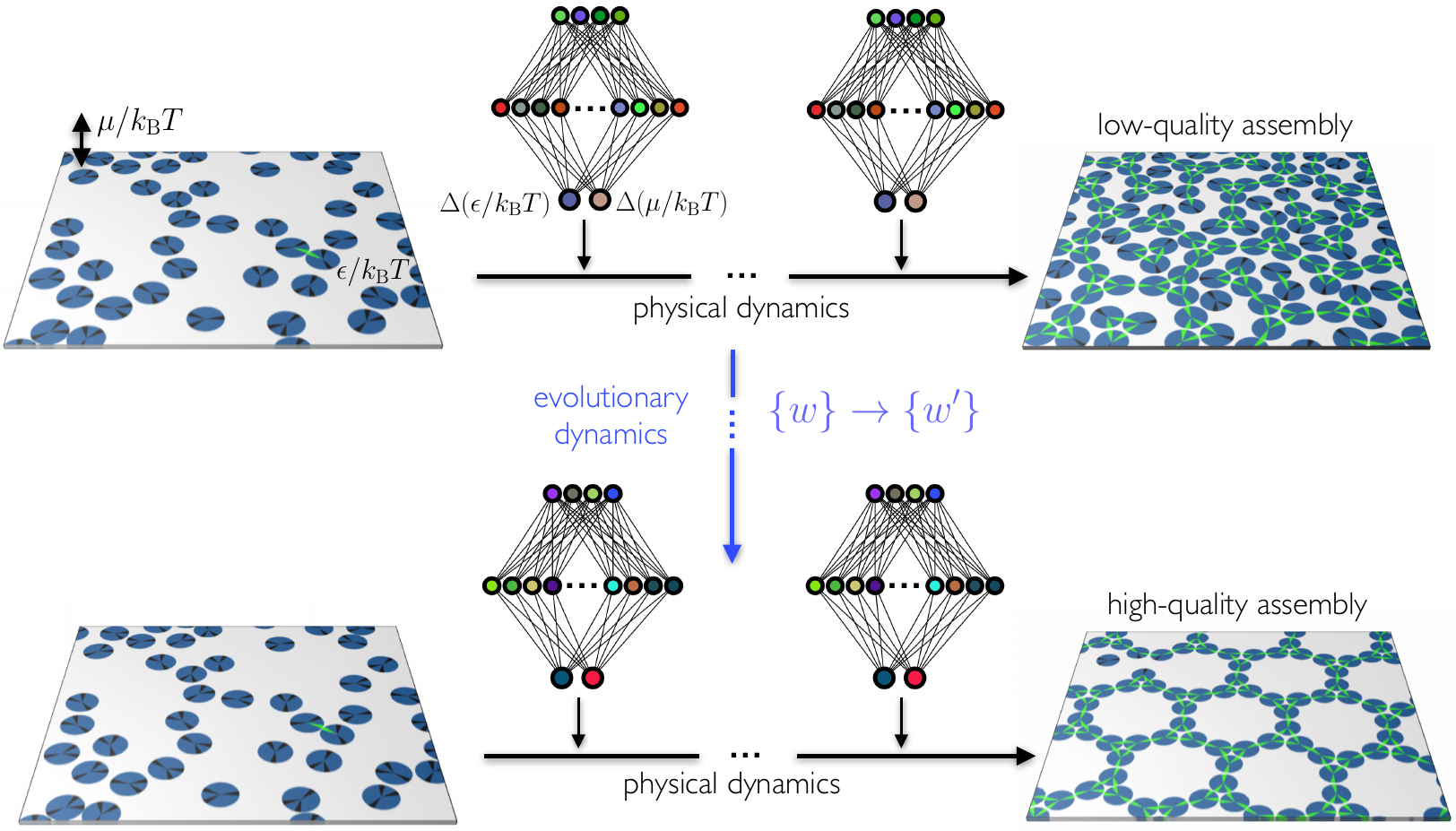} 
   \caption{\label{fig1} In this paper we show that neural-network policies trained by evolutionary reinforcement learning can enact efficient time- and configuration-dependent protocols for molecular self-assembly. A neural network periodically controls certain parameters of a system, and evolutionary learning applied to the weights of a neural network (indicated as colored nodes) results in networks able to promote the self-assembly of desired structures. The protocols that give rise to these structures are then encoded in the weights of a self-assembly kinetic yield net.}
\end{figure*}

Absent a theoretical framework for self-assembly, an alternative is to seek assistance from machine learning in order to attempt to control self-assembly without human intervention. In this paper we show that neural-network-based evolutionary reinforcement learning can be used to develop protocols for the control of self-assembly, without prior understanding of what constitutes a good assembly protocol. Reinforcement learning is a branch of machine learning concerned with learning to perform actions so as to achieve an objective\c{sutton2018reinforcement}, and has been used recently to play computer games better than humans can\c{QL,DQN,Atari,Atari2600,Actor,DeepMind,MuJoCo,MDP,Rogue,NFQ,Soccer,PPO,OpenAI,VizDoom,VizDoom2,Go,Go2,Guber}. Neuroevolution\c{GA,GA2,lehman2018more,salimans2017evolution,zhang2017relationship,lehman2018safe,conti2018improving,Guber} is an approach to reinforcement learning that is much less widely applied than value-based methods\c{sutton2018reinforcement}, but is a simple and powerful method that is naturally suited to ``sparse-reward'' problems such as self-assembly, where the outcome of assembly (good or bad) is not always apparent until its latter stages. Here we apply neuroevolutionary learning to stochastic molecular simulations of patchy particles, a standard choice for representing anisotropically-interacting molecules, nanoparticles, or colloids\c{zhang2004self,romano2011crystallization,sciortino2007self,doye2007controlling,bianchi2008theoretical,doppelbauer2010self,whitelam2014common,duguet2016patchy}. While a neural network cannot influence the fundamental dynamical laws by which such systems evolve\c{frenkel1996understanding}, it can control the parameters that appear in the dynamical algorithm, such as temperature, chemical potential, and other environmental conditions. In this way the network can influence the sequence of microstates visited by the system. We show that a neural network can learn to enact a time-dependent protocol of temperature and chemical potential (called a {\em policy} in reinforcement learning) in order to promote the self-assembly of a desired structure, or choose between two competing polymorphs. In both cases the network identifies strategies different to those informed by human intuition, but which can be analyzed and used to provide new insight. We use networks that take only elapsed time as their input, and networks that take microscopic information from the system. Both learn under evolution, and networks permitted microscopic information learn better than those that are not. 

Networks enact protocols that are out of equilibrium, in some cases far from equilibrium, and so are not well-described by existing theories. These ``self-assembly kinetic yield'' networks act to promote a particular order parameter for self-assembly at the end of a given time interval, with no consideration for whether a process results in an equilibrium outcome or not. It is therefore distinct from feedback approaches designed to promote near-equilibrium behavior\c{klotsa2013controlling}. Our approach is similar in intent to \cc{tang2016optimal}, in which dynamic programming is used to find protocols able to promote colloidal crystallization using an external field. One important difference between that work and ours is that our scheme does not require measurement of the order parameter we wish to promote (except at the end of the experiment), making it applicable to molecular and nanoscale systems whose microscopic states cannot be observed as they evolve. We also use a neural network to encode the assembly protocol, rather than a model of discretized states. Our approach is also similar to that of \cc{miskin2016turning} in the sense that we use evolutionary search to optimize assembly, but we allow the learning procedure to respond to both temporal and microscopic information via the use of a neural network. Our approach is complementary to efforts that use machine learning to analyze existing self-assembly pathways\c{long2015machine,long2014nonlinear}, or to infer or design structure-property relationships for self-assembling molecules\c{lindquist2016communication,thurston2018machine,ferguson2017machine}. The present scheme is simple and can be straightforwardly altered to observe an arbitrary number of system features, and to control an arbitrary number of system parameters, and so can be applied to a wide range of experimental systems.

In \s{sec_ev} we describe the evolutionary scheme, which involves alternating physical and evolutionary dynamics. In \s{sec_ass} we show that it leads to networks able to promote the self-assembly of a certain structure faster and better than intuitive cooling protocols can. In \s{sec_poly} we show that networks can learn to select between two polymorphs that are equal in energy and that form in unpredictable quantities under slow cooling protocols. The strategy used by the networks to achieve this selection provides new insight into the self-assembly of the system under study. We conclude in \s{sec_conc}. Networks learn these efficient and new self-assembly protocols with no human input beyond the specification of which target parameter to promote, pointing the way to the design of synthesis protocols by artificial intelligence.

\section{Evolutionary reinforcement learning of self-assembly protocols}
\label{sec_ev}

We sketch in \f{fig1} an evolutionary scheme by which a self-assembly kinetic yield net can learn to control self-assembly. We consider a computational model of molecular self-assembly, patchy discs of diameter $a$ on a two-dimensional square substrate of edge length $50a$. The substrate (simulation box) possesses periodic boundary conditions in both directions. Discs, which cannot overlap, are minimal representations of molecules, and the patches denote their ability to make mutual bonds at certain angles. By choosing certain patch angles, widths, and binding-energy scales it is possible to reproduce the dynamic and thermodynamic behavior of real molecular systems of a broad range of lengthscales and material types\c{whitelam2014common}. The disc model is a good system on which to test the application of evolutionary learning to self-assembly, because it is simple enough to simulate for long times, and its behavior is complex enough to capture several aspects of real self-assembly, including the formation of competing polymorphs and structures that are not the thermodynamically stable one. Choosing protocols to promote the formation of particular structures within the disc model is therefore nontrivial, and serves as a meaningful test of the learning procedure.

Two discs receive an energetic reward of $-\epsilon/\kt$ if their center-to-center distance $r$ is between  $a$ and $a+a/10$, and if the line joining those discs cuts through one patch on each disc\c{kern2003fluid}. In addition, we sometimes require patches to possess certain identities in order to bind, mimicking the ability of e.g. DNA to be chemically specific\c{whitelam2016minimal}. In this paper we consider disc types with and without DNA-type specificity. Bound patches are shown green in figures, and unbound patches are shown black. In figures we often draw the convex polygons formed by joining the centers of bound particles\c{whitelam2014common}. Doing so makes it easier to spot regions of order by eye. Polygon counts serve as a useful order parameter for self-assembly, because they are related (in some cases proportional) to the number of unit cells of the desired material. We denote by $N_\alpha$ the number of convex $\alpha$-gons within a simulation box.

We simulated this system in order to mimic an experiment in which molecules are deposited on a surface and allowed to evolve. We use two stochastic Monte Carlo algorithms to do so. One is a grand-canonical algorithm that allows discs to appear on the substrate or disappear into a notional solution\c{frenkel1996understanding}; the other is the virtual-move Monte Carlo algorithm\c{whitelam2009role, VMMC_3} that allows discs to move collectively on the surface in an approximation of Brownian motion\c{haxton2015crystallization}. If $M$ is the instantaneous number of discs on the surface then we attempt virtual moves with probability $M/(1+M)$, and attempt grand-canonical moves otherwise. Doing so ensures that particle deposition occurs at a rate (for fixed control parameters) that is roughly insensitive of substrate density. The acceptance rates for grand-canonical moves are given in \cc{whitelam2014common} (essentially the textbook rates\c{frenkel1996understanding} with the replacement $M \to M+1$ to preserve detailed balance in the face of a fluctuating proposal rate). One such decision constitutes one Monte Carlo step~\footnote{The natural way to measure ``real'' time in such a system is to advance the clock by an amount $(1+M)^{-1}$ upon making an attempted move. Dense systems and sparse systems then take very different amounts of CPU time to run. In order to move simulation generations efficiently through our computer cluster we instead updated the clock by one unit upon making a move. In this way we work in the constant event-number ensemble.}.

The grand-canonical algorithm is characterized by a chemical potential $\mu/\kt$, where $\kt$ is the energy scale of thermal fluctuations. Positive values of this parameter favor a crowded substate, while negative values favor a sparsely occupied substrate. If the interparticle bond strength $\epsilon/\kt$ is large, then there is, in addition, a thermodynamic driving force for particles to assemble into structures. (In experiment, bond strength can be controlled by different mechanisms, depending upon the physical system, including temperature or salt concentration; here, for convenience, we sometimes describe increasing $\epsilon/\kt$ as ``cooling'', and decreasing $\epsilon/\kt$ as ``heating''.) For fixed values of these parameters the simulation algorithm obeys detailed balance, and so the system will evolve toward its themodynamic equilibrium. Depending on the parameter choices, this equilibrium may correspond to an assembled structure or to a gas or liquid of loosely-associated discs. For finite simulation time there is no guarantee that we will reach this equilibrium. Here we consider evolutionary simulations or trajectories of $t_0=10^9$ individual Monte Carlo steps (not sweeps, or steps per particle), starting from substrates containing 500 randomly-placed non-overlapping discs. These are relatively short trajectories in self-assembly terms: the slow cooling protocols of \cc{whitelam2016minimal} used trajectories about 100 times longer. 

Each trajectory starts with control-parameter values $\epsilon/\kt=3$ and $\mu/\kt=2$, which does not give rise to self-assembly. As a trajectory progresses, a neural network chooses, every $10^{-3} t_0$ Monte Carlo steps, a change $\Delta(\mu/\kt)$ and $\Delta(\epsilon/\kt)$ of the two control parameters of the system (and so the same network acts 1000 times within each trajectory). These changes are added to the current values of the relevant control parameter, as long as they remain within the intervals $\epsilon/\kt \in [0,20]$ and $\mu/\kt \in [-20,20]$ (if a control parameter moves outside of its specified interval then it is returned to the edge of the interval). Between neural-network actions, the values of the control parameters are held fixed. Networks are fully-connected architectures with 1000 hidden nodes and two output nodes, and a number of input nodes appropriate for the information they are fed. We used tanh activations on the hidden nodes; the full network function is given in \s{sec_net}.

Training of the network is done by evolution\c{Guber}. We run 50 initial trajectories, each with a different, randomly-initialized neural network. Each network's weights and biases $\{w\}$ are independent Gaussian random numbers of zero mean and unit variance. The collection of 50 trajectories produced by this set of 50 networks is called generation 0. After these trajectories run we assess each according to the number $N_{\alpha}$ of convex $\alpha$-gons present in the simulation box; the value of $\alpha$ depends on the disc type under study and the structure whose assembly we wish to promote. The 5 networks whose trajectories have the largest values of $N_{\alpha}$ are chosen to be the ``parents'' of generation 1. Generation 1 consists of these 5 networks, plus 45 mutants. Mutants are made by choosing at random one of the parent networks and adding to each weight and bias a Gaussian random number of zero mean and variance 0.01. After simulation of generation 1 we repeat the evolutionary procedure in order to create generation 2. Alternating the physical dynamics (the self-assembly trajectories) and the evolutionary dynamics (the neural-network weight mutation procedure) results in populations of networks designed to control self-assembly conditions so as to promote certain order parameters.

Each evolutionary scheme used one of three types of network. The first, called the {\em time network} for convenience, has a single input node that takes the value of the scaled elapsed time of the trajectory, $t/t_0 \in [0,1]$. The second, called the {\em microscopic network} for convenience, has $P+1$ input nodes, where $P$ is the number of patches on the disc. Input node $i \in \{0,1,...,P\}$ takes the value $S_i$, the number of particles in the simulation box that possess $i$ engaged patches (divided by 1000). The third neural network type has $P+2$ input nodes, and takes both $t/t_0$ and the $S_i$ as inputs. We chose the time network so as to explore the ability of a network to influence the self-assembly protocol if it cannot observe the system at all. We chose the microscopic network to see if a network able to observe the system can do better than one that cannot. We do not intend for its input to be a precise analog of an experimental measurement, but there are several experimental techniques able to access similar information, such as the averaged number of particles in certain types of environment, or the approximate degree of aggregation present in a system\c{de2015crystallization}.

It is important to note that these microscopic inputs are not related in a simple way to the evolutionary parameters $N_{\alpha}$, the number of convex $\alpha$-gons in the box, that we wish to optimize. For instance, in \s{sec_ass}, both dense disordered networks (with small values of $N_{12}$) and well-assembled structures (with large values of $N_{12}$) can contain similar numbers of maximally-coordinated particles. In \s{sec_poly}, the two polymorphs we ask a network to choose between, one described by $N_6$ and the other by $N_4$, have identical coordination numbers. Thus a network must learn the connection between the data it is fed and the evolutionary order parameters we aim to maximize. Our intent was to mimic an experiment in which which some microscopic information about a system is available, but the quality of assembly can only be assessed after the experiment has run to completion. The success of the learning scheme in the absence of any system-specific information, and our finding that the more information we feed a network the better it performs, suggests that the evolutionary scheme can be applied to a wide variety of experimental systems.

Dynamical trajectories are stochastic, even given a fixed protocol (policy), and so networks that perform well in one generation may be eliminated the next. This can happen if, for example, a certain protocol promotes nucleation, the onset time for which varies from one trajectory to another. By the same token, the best yield can decrease from one generation to the next, and independent trajectories generated using a given protocol have a range of yields. To account for this effect one could place evolutionary requirements on the yield associated with many independent trajectories using the same protocol. Here we opted not to do this, reasoning that over the course of several generations the evolutionary process will naturally identify protocols that perform well when measured over many independent trajectories. We demonstrate this feature in \s{sec_ass}, where independent trajectories produced under slow cooling display a wide variety of outcomes, but independent trajectories generated by evolved protocols display relatively well-defined ones.

\section{Promoting self-assembly}
\label{sec_ass}
\begin{figure}[] 
   \centering
\includegraphics[width=\linewidth]{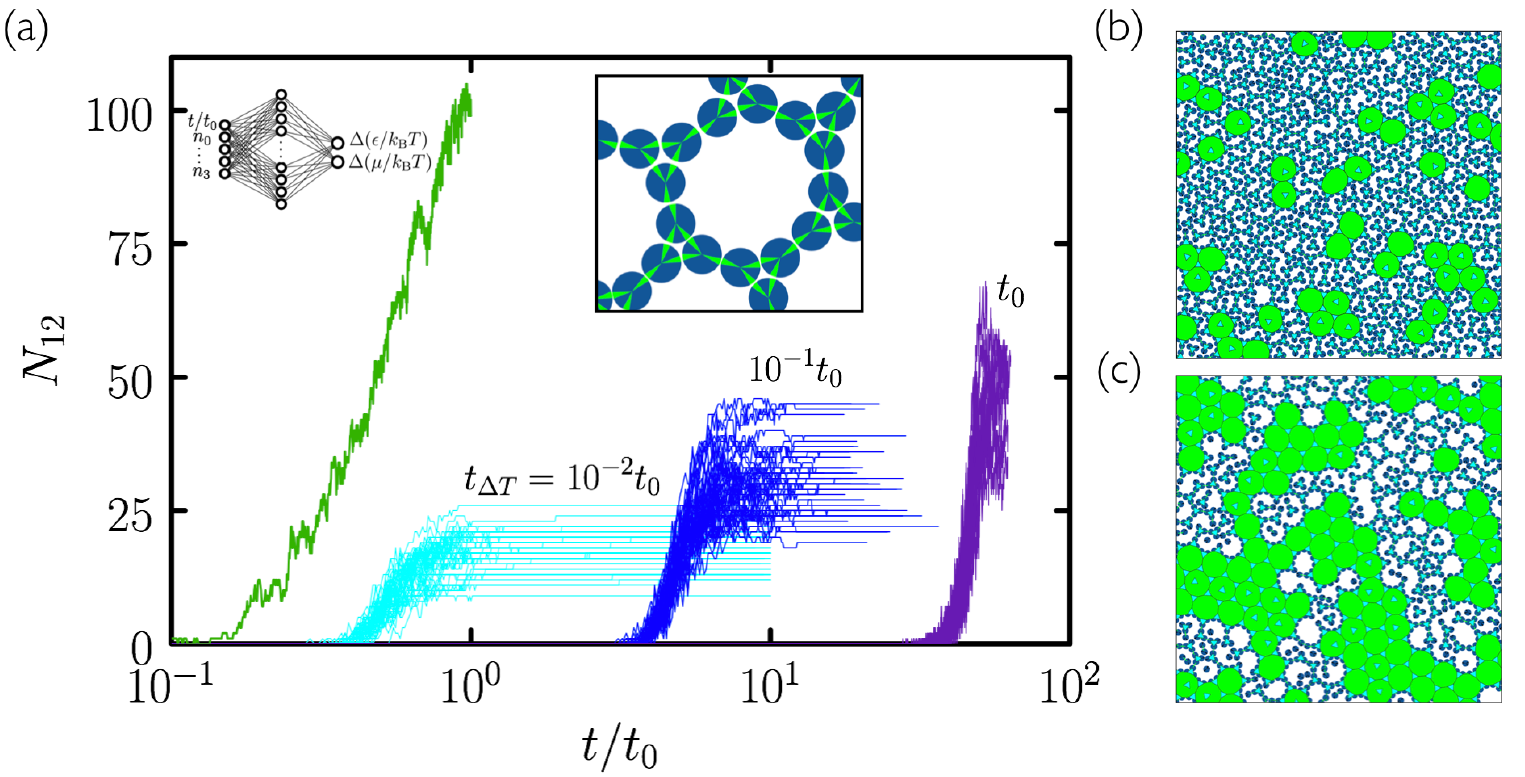} 
 \caption{\label{fig_cool}  (a) A 3-patch disc with chemically selective patches can form a structure equivalent to the 3.12.12 Archimedean tiling\c{whitelam2016minimal}, a tiling with one 3-gon and two 12-gons around each vertex (inset). Slow cooling simulations, in which the disc interaction strength $\epsilon/\kt$ is increased by 0.075 every $t_{\Delta T}$ Monte Carlo steps, give rise to the numbers of 12-gons $N_{12}$ (the number of unit cells of the desired structure) shown in the plot: we show 50 independent trajectories at each of three cooling rates. Neural networks learn to control $\epsilon/\kt$ and $\mu/\kt$ in order to greatly exceed these yields, in a fraction of the time (green line at left). Panels (b) and (c) show snapshots of structures produced by slow cooling and a neural-network protocol, respectively, with 12-gons colored green.}
\end{figure}

\begin{figure*}[] 
   \centering
\includegraphics[width=\linewidth]{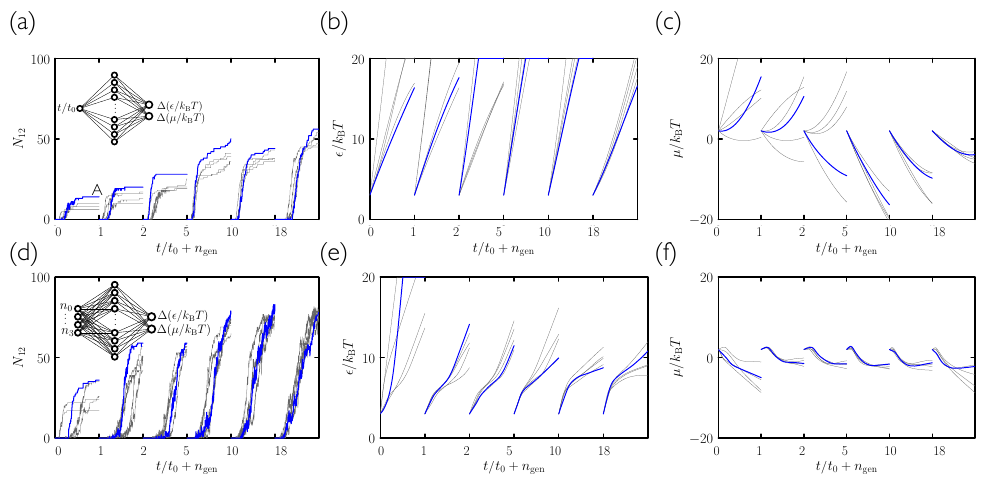} 
 \caption{\label{fig2} Evolutionary learning of self-assembly protocols using the 3.12.12 disc of \f{fig_cool}. (a) The time network used with this disc, within the evolutionary scheme of \f{fig1}, produces progressively better yields of 12-gons with generation. We show the top 5 yields per generation, with the best shown in blue (dark gray). The protocols leading to these yields are shown in (b,c), the better yields corresponding to rapid cooling and evacuation of the substrate. (d) The microscopic network used in the same evolutionary scheme produces better yields than the time network, using (e,f) similar but slightly more nuanced protocols. Networks that take both temporal and microscopic information, or two networks used in sequence, produce better yields still: see \f{fig_protocols}.}
\end{figure*}

\begin{figure}[] 
   \centering
\includegraphics[width=\linewidth]{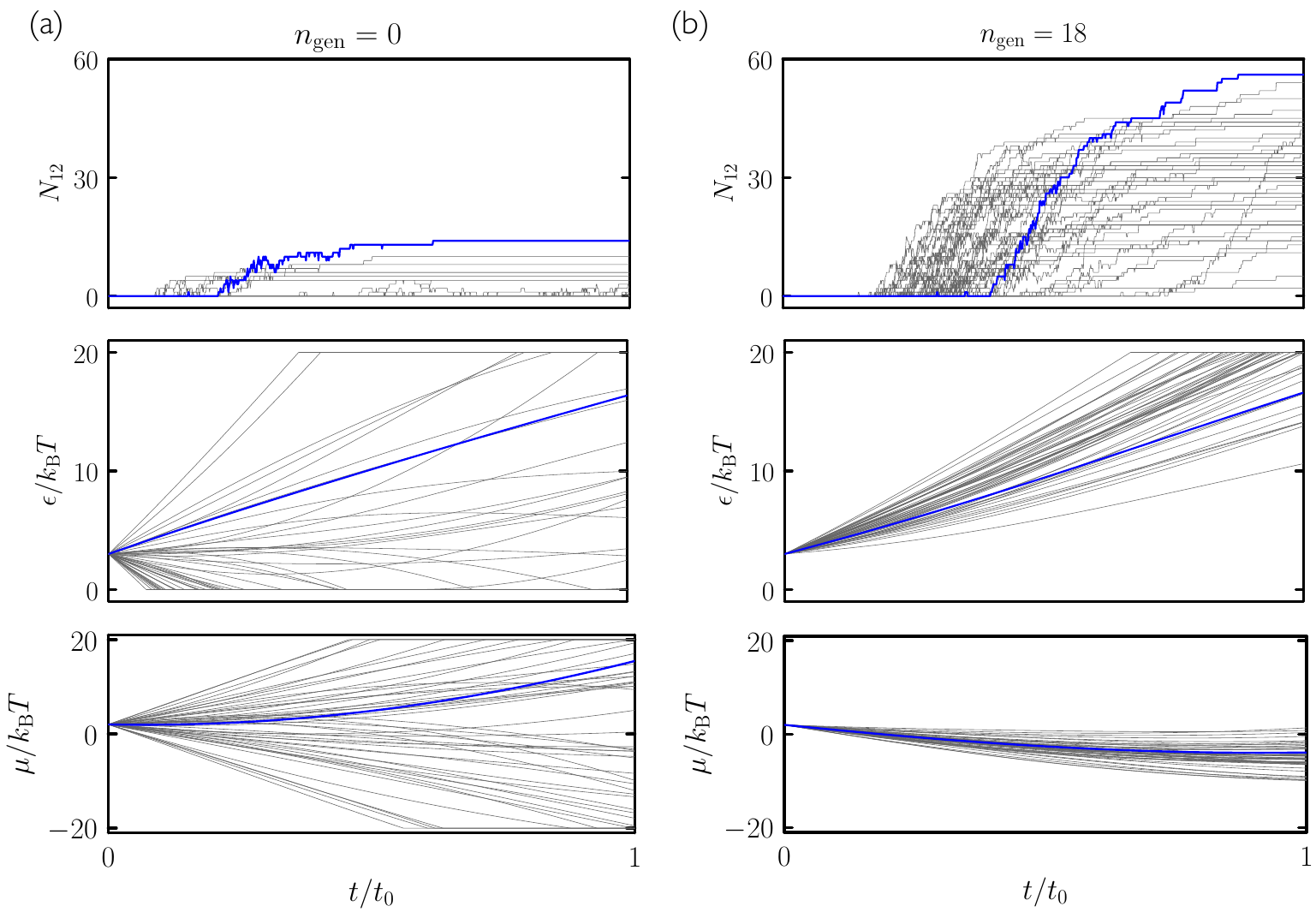} 
   \caption{\label{fig2_supp} (a) Generation-0 trajectories of the time network applied to the 3.12.12 disc; most networks fail to produce assembly. (b) Generation-18 trajectories generally result in much better assembly. However, note that some networks, although they are offspring of successful generation-17 networks, result in low-quality assembly. }
\end{figure}

\begin{figure}[] 
   \centering
\includegraphics[width=\linewidth]{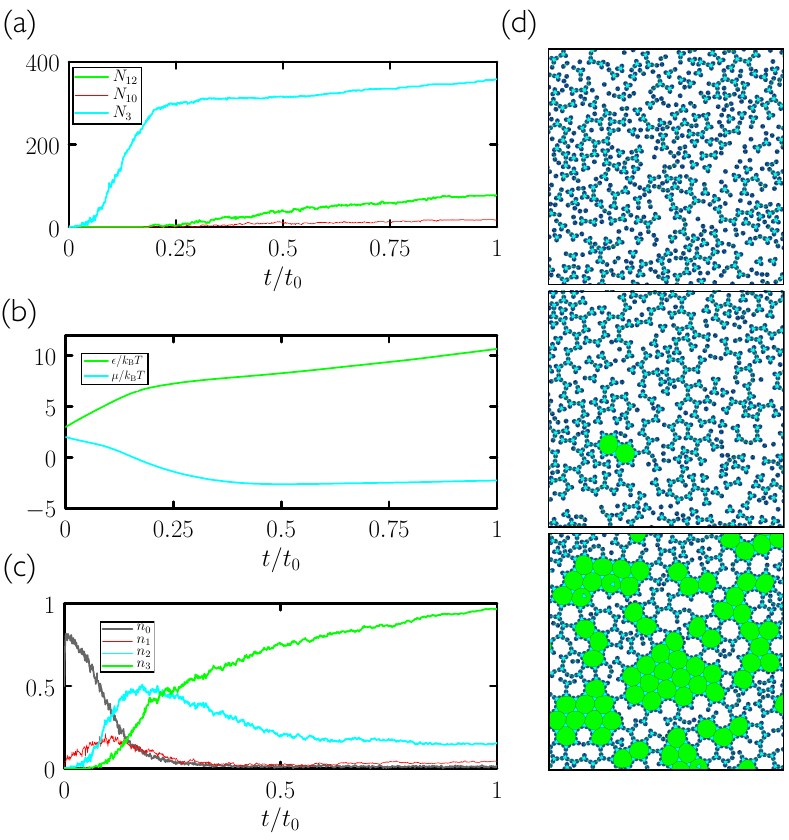} 
   \caption{\label{fig3} A self-assembly trajectory produced by the best generation-18 microscopic network of \f{fig2}(d--f). Panel (a) and the time-ordered snapshots in (d) show the dynamics to be hierarchical in an extreme way, with most 3-gons (blue) forming before 12-gons (green) are made. Snapshot times are $t/t_0=0.17,0.25,1$, from top to bottom. More detail can be seen in snapshots by enlarging them on a computer screen. Defects, such as disordered regions and 10- and 14-gons, also form. Panel (b) shows the temperature and chemical potential protocols chosen by the network, and (c) shows the inputs to the network.}
\end{figure}

\begin{figure}[] 
   \centering
\includegraphics[width=0.9\linewidth]{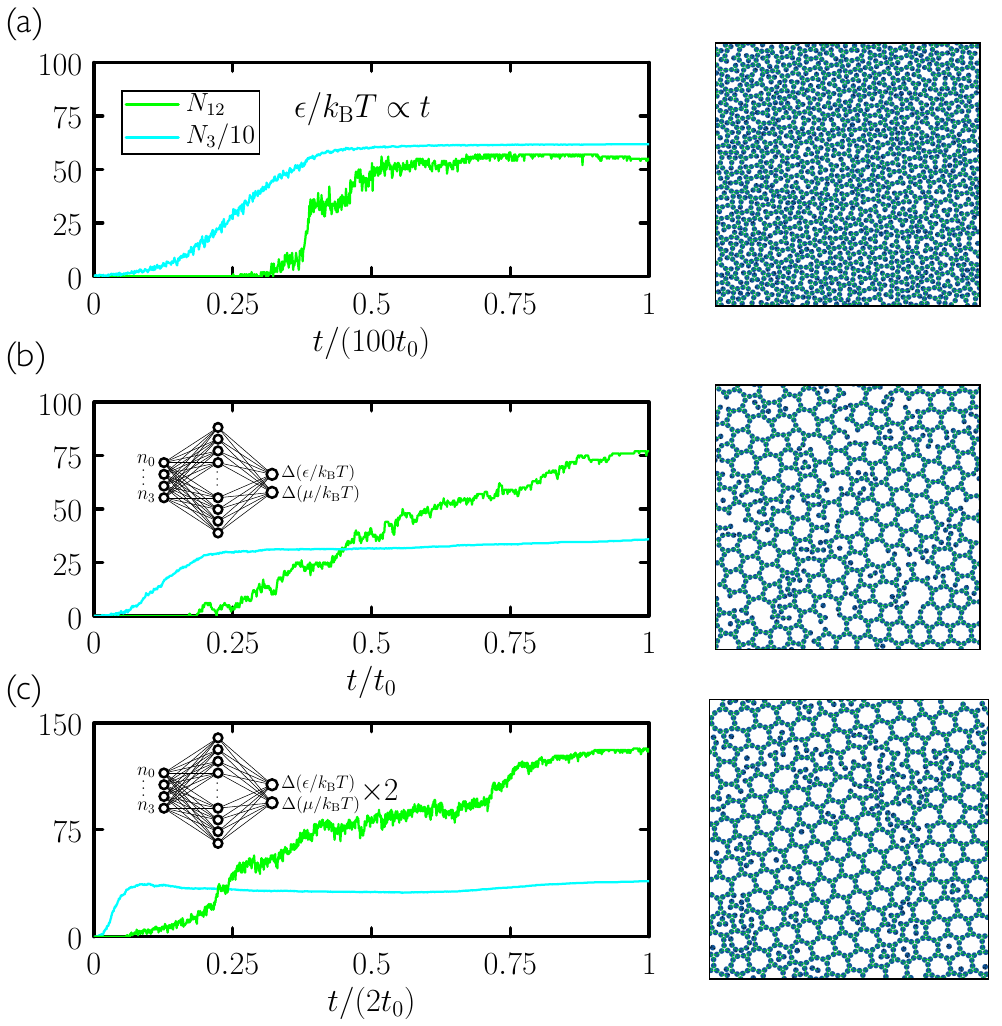} 
   \caption{\label{fig2_supp2} We compare trajectories produced by (a) the slowest cooling rate shown in \f{fig_cool}, (b) the best generation-18 microscopic network from \f{fig2}(d--f), and (c) a generation-12 procedure using two microscopic networks in sequence; see \f{fig_protocols}(e--h). The neural networks produce better assembly than the cooling protocol (measured by the 12-gon count, i.e. the number of unit cells of the desired structure), and do so 50 or 100 times faster. The snapshots at right are taken at the end of the three trajectories. More detail can be seen in snapshots by enlarging them on a computer screen. In (a), the 3.12.12 structure contains many smaller species in its pores. Note also that some of the larger closed loops in these images are 10-gons or 14-gons; the polygon representation of \f{fig3} picks out 12-gons more clearly.}
\end{figure}

\begin{figure*}[] 
   \centering
\includegraphics[width=\linewidth]{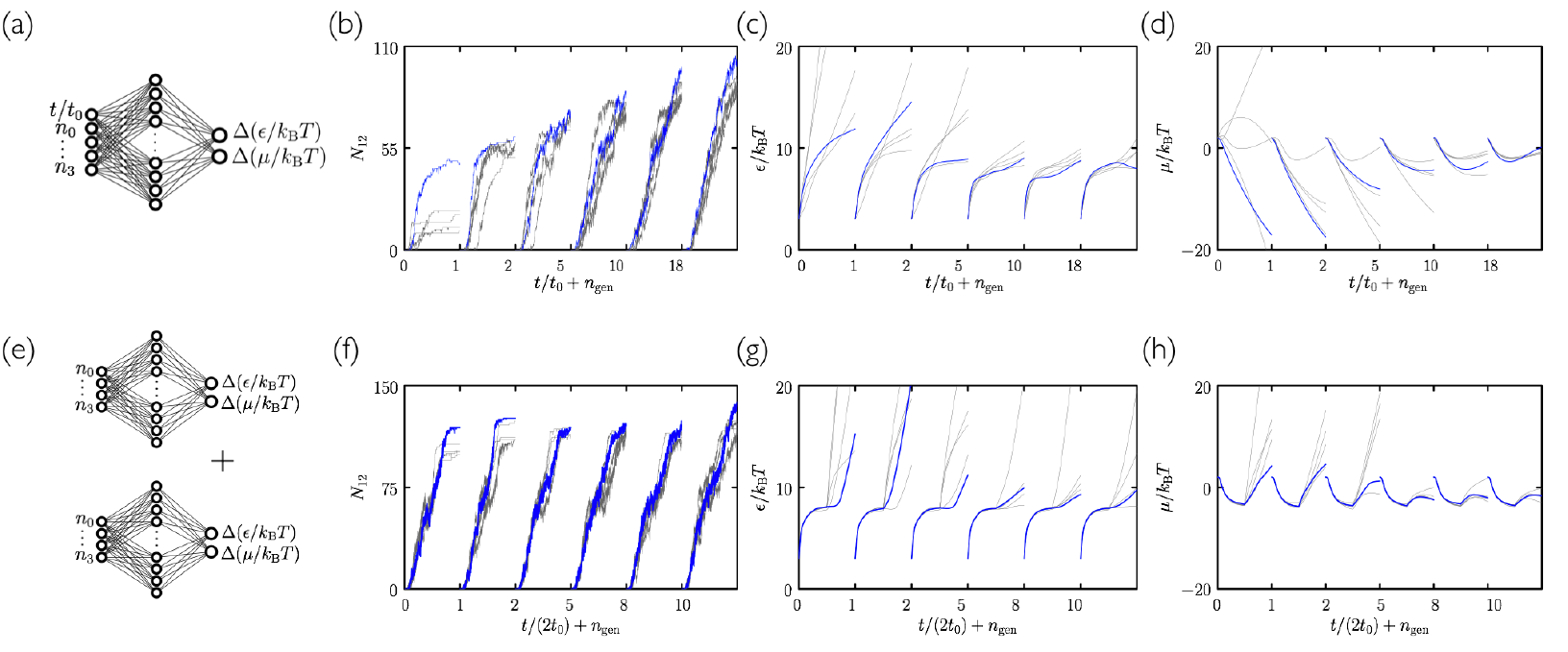} 
   \caption{\label{fig_protocols} (a) A neural network that combines temporal and microscopic information outperforms both the time- and microscopic networks of \f{fig2}. Panel (b) shows the yield of the top 5 of 50 trajectories for certain generations; panels (c) and (d) show the associated values of $\epsilon/\kt$ and $\mu/\kt$, respectively. The protocol learned by this network is similar to that leaned in \f{fig2}, but with more pronounced non-monotonicity: at later times the substrate is heated and made more dense, which appears to facilitate annealing of the structures grown under cold, sparse conditions. A generation-18 trajectory of this network is shown in \f{fig_cool}(c) (at left). In panels (e--h) we show results using a two-step procedure: the best generation-18 microscopic network from \f{fig2} is applied for time $t_0$, and then a second microscopic network is applied and trained for a period $t_0$. This procedure identifies a protocol similar to that shown in the upper panels, whose early and late stages are suggestive of distinct growth and annealing conditions. These strategies produce yield of order twice that obtained under slow cooling; see \f{fig_cool} and \f{fig2_supp2}.}
\end{figure*}

In \f{fig2} we consider the ``3.12.12'' disc of \cc{whitelam2016minimal}, which has three, chemically specific patches whose bisectors are separated by angles $\pi/3$ and $5\pi/6$. This disc can form a structure equivalent to the 3.12.12 Archimedean tiling (a tiling with one 3-gon and two 12-gons around each vertex). The number of 12-gons $N_{12}$ counts the number of unit cells of the structure, and so is a suitable order parameter for evolutionary search. This structure is a difficult target for self-assembly because its unit cell is large and must form from floppy intermediates, the nature of which gives plenty of scope for mistakes of binding and kinetic trapping. As a result, while intuitive protocols allow assembly to proceed, they do so with relatively low fidelity. In \f{fig_cool}(a) we show the outcome of ``cooling'' simulations done at three different rates. As for evolutionary simulations, we start from control-parameter values $\epsilon/\kt=3$ and $\mu/\kt=2$, where the equilibrium state is a sparse gas of largely unassociated discs. Every $t_{\Delta T}$ Monte Carlo steps we increase $\epsilon/\kt$ by a value 0.075. We carried out 50 independent simulations at each cooling rate. As the rate of cooling decreases, the yield increases, but achieving much more than 50 unit cells of the target material is time-consuming: single trajectories at each of the cooling rates take, respectively, of order an hour, a day, and a week of CPU time on a single processor. Clearly, substantial improvement using this protocol would require prohibitively long simulations.

Search using evolutionary learning results in protocols that can greatly exceed the yield of cooling simulations, in a fraction of the time (an example is shown at left in \f{fig_cool}(a)). In \f{fig2}(a--c) we show results obtained using the time network within the evolutionary scheme of \f{fig1}. Generation-0 trajectories are controlled by essentially random protocols, and many (e.g. those that involve weakening of interparticle bonds) result in no assembly (see \f{fig2_supp}). Some protocols result in low-quality assembly (comparable to that seen in the fastest cooling protocols of \f{fig_cool}), and the best of these are used to create generation 1. \f{fig2}(a) shows that assembly gets better with generation number: evolved networks learn to promote assembly of the desired structure. The protocols leading to these structures are shown in \f{fig2}(b,c): early-generation networks tend to strengthen bonds (``cool'') quickly and concentrate the substrate, while later-generation networks strengthen bonds {\em more} quickly but also promote evacuation of the substrate. This strategy appears to reduce the number of obstacles to the closing of the large and floppy intermediate structures. The most advanced networks further refine these bond-strengthening and substrate-evacuation protocols.

The microscopic network [\f{fig2}{(d--f)] produces slightly more nuanced versions of the time-network protocols, and leads to better assembly. Thus, networks given access to configurational information learn more completely than those that know only the elapsed time of the procedure, even though the information they are given does not directly relate to the quality of assembly. In \f{fig3} we show in more detail a trajectory produced by the best generation-18 microscopic network. The self-assembly dynamics that results is hierarchical assembly of the type seen in \cc{whitelam2016minimal}, in which trimers (3-gons) form first, and networks of trimers then form 12-gons, but is a more extreme version: in \f{fig3} we see that almost all the 3-gons made by the system form before the 12-gons begin to form. Thus the network has adopted a two-stage procedure in an attempt to maximize yield.  

Networks given either temporal or microscopic information have therefore learned to promote self-assembly, without any external direction beyond an assessment, at the end of the trajectory, of which outcomes were best. Moreover, the quality of assembly considerably exceeds the quality of intuition-driven cooling procedures, and proceeds much more quickly. In \f{fig2_supp2} we compare trajectories and assembled structures produced by cooling and by two different networks: the networks produce better structures, even though they are constrained to act over much shorter times. Here we observe the counterintuitive result of rapidly-varying nonequilibrium protocols producing better-quality assembly than a slow-cooling procedure designed (at least in an intuitive sense) to promote ``near equilibrium'' conditions\c{whitelam2018strong}.

Yield under the evolutionary protocols can be increased by providing more data to the neural network. In \f{fig_protocols} we show that a neural network provided with both temporal and microscopic information outperforms both the time- and the microscopic networks of \f{fig2}. Yield can also be increased by using two neural networks, one after the other, trained independently [see \f{fig2_supp2}(c) and \f{fig_protocols}(e--h)]. In these cases the yield of material reaches more than double that obtained under a slow-cooling protocol. Protocols learned by these neural networks are distinctly different at early and late times, suggestive of distinct growth and annealing stages: after an initial stage of rapid growth under cool and sparse conditions, networks heat the substrate and make it more dense, apparently in order to promote error-correction.

Here we have provided no prior input to the neural net to indicate what constitutes a good assembly protocol. One could alternatively survey parameter space as thoroughly as possible, using intuition and prior experience, before turning to evolution. In such cases generation-0 assembly would be better than under randomized protocols. However, we found that even when generation-0 assembly was already of high quality, the evolutionary procedure was able to improve it. In \f{fig2_supp3} we consider evolutionary learning using the regular three-patch disc without patch-type specificity\c{whitelam2014common,whitelam2016minimal}. This disc forms the honeycomb network so readily that the best examples of assembly using 50 randomly-chosen protocols (generation 0) are already good. Nonetheless, evolution using the time network or microscopic network is able to improve the quality of assembly, with the microscopic network again performing better.

\begin{figure}[] 
   \centering
\includegraphics[width=\linewidth]{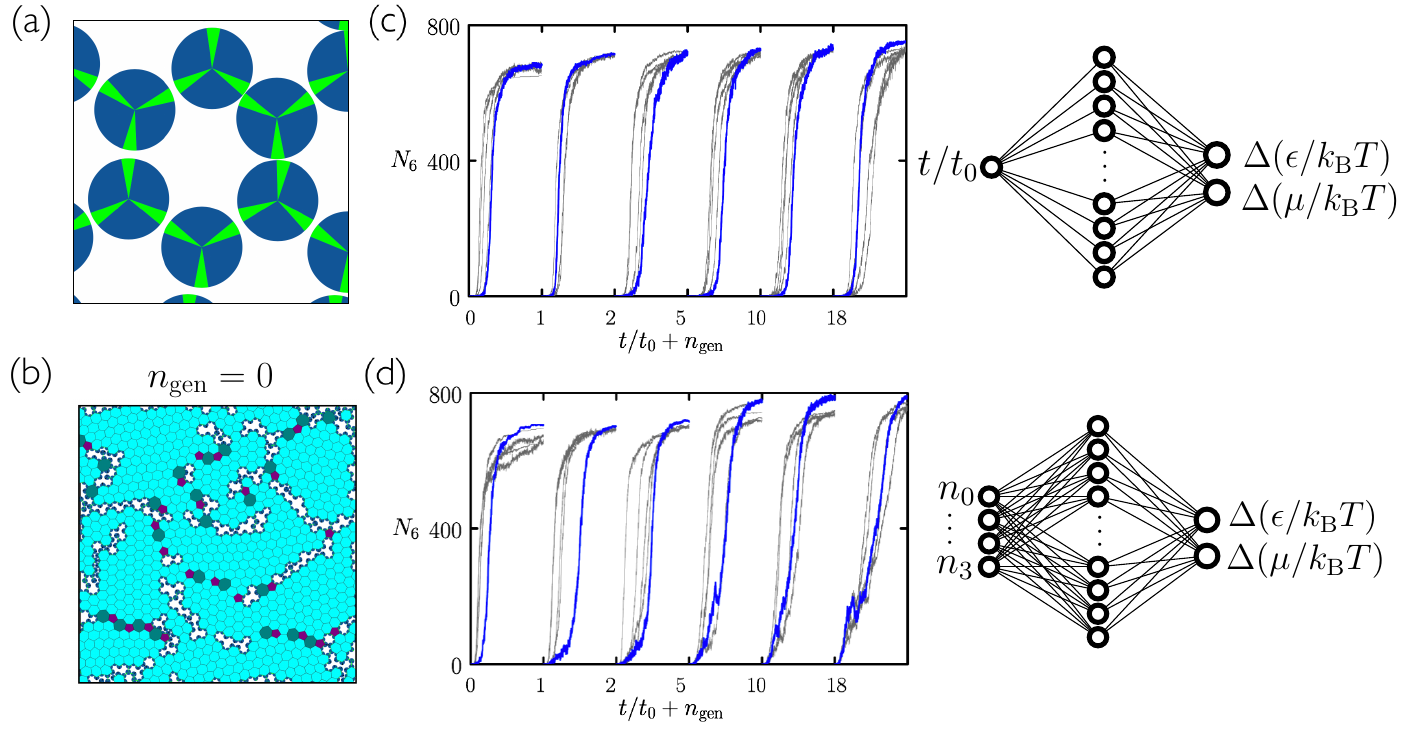} 
   \caption{\label{fig2_supp3} Evolutionary learning of self-assembly protocols with the regular three-patch disc without patch specificity. This disc forms the honeycomb network (a) so readily that assembly using 50 randomly-chosen protocols (generation-0) is already good; see panel (b), in which 6-gons are colored light blue (gray). Nonetheless, evolution using the time network (c) or microscopic network (d) can improve the quality of assembly, and, again, the microscopic network is better than the time one. We show the top 5 trajectories per generation, with the best shown in blue (dark gray).}
\end{figure}

\section{Polymorph selection}
\label{sec_poly}
\begin{figure}[] 
   \centering
\includegraphics[width=\linewidth]{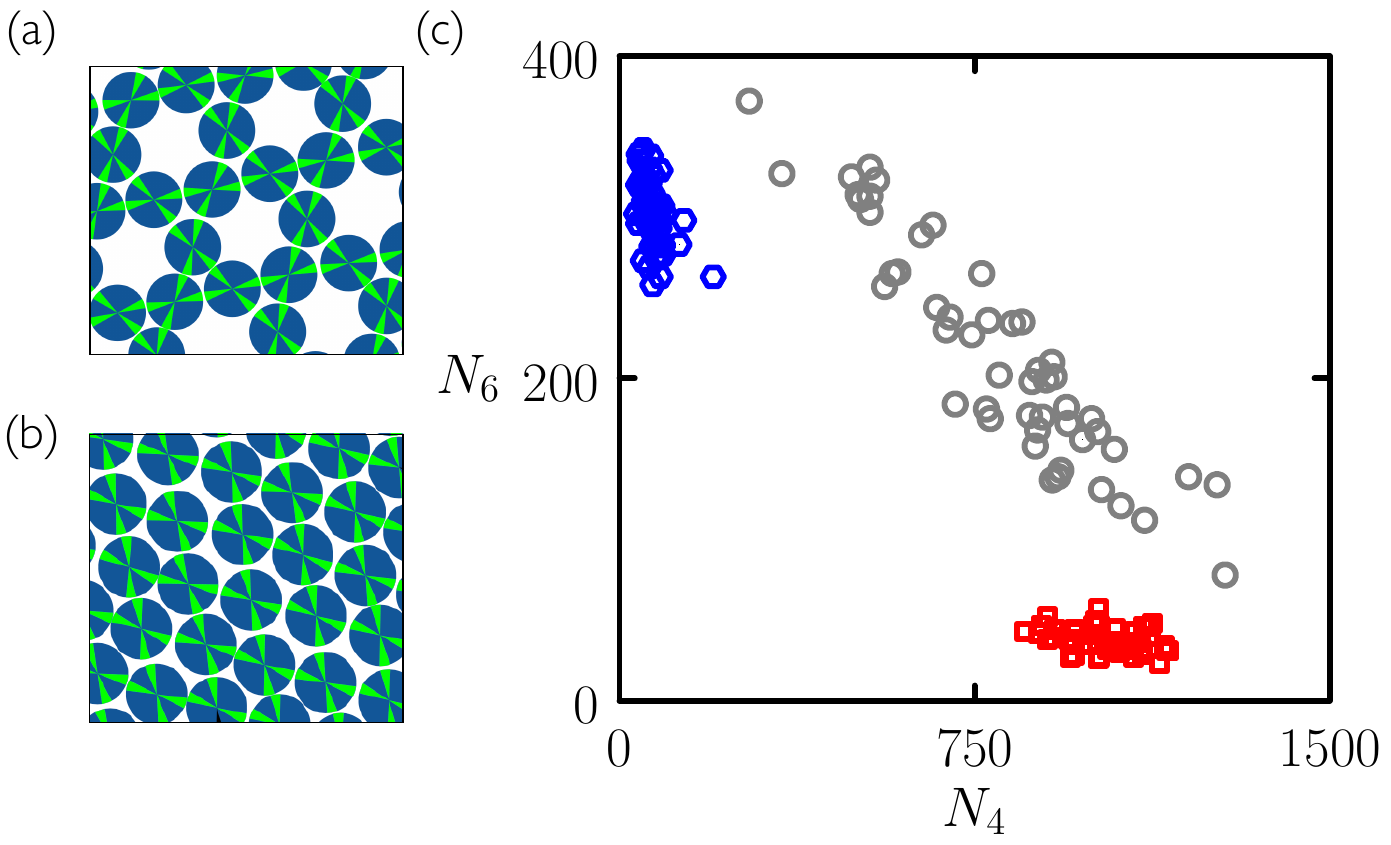} 
 \caption{\label{fig_poly}  A 4-patch disc with angles $\pi/3$ and $2 \pi/3$ between patch bisectors can form (a) a structure equivalent to the 3.6.3.6 Archimedean tiling or (b) a rhombic structure\c{whitelam2016minimal}. The number of 6-gons $(N_6)$ or 4-gons $(N_4)$ serve as order parameters for these structures. (c) Cooling (at the slowest rate shown in \f{fig_cool}) causes nucleation and growth of the two polymorphs on a timescale of order $20 t_0$ (see \f{fig_mech}). The outcome of 50 such trajectories consists of either or both polymorphs, in an unpredictable way (gray circles). By contrast, the evolutionary scheme of \f{fig1} can produce neural networks able to select either polymorph with high fidelity. Blue (dark grey) hexagons (resp. red (light gray) squares) correspond to 50 trajectories, of length $t_0$, using a single generation-10 microscopic neural network evolved so as to maximize $N_6$ (resp. $N_4$); see \f{fig5}.}
\end{figure}

 \begin{figure*}[] 
   \centering
\includegraphics[width=\linewidth]{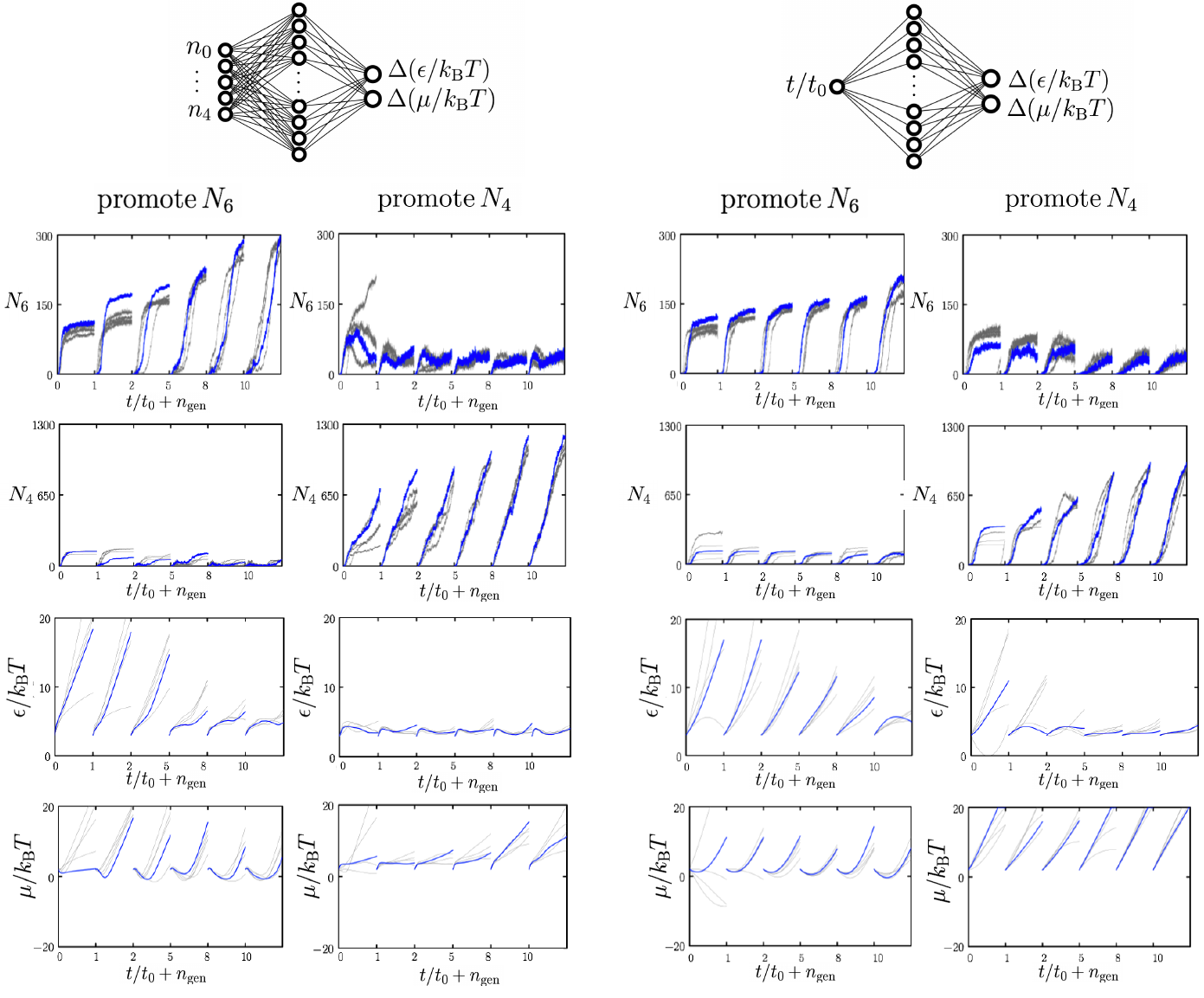} 
   \caption{\label{fig5} Evolutionary learning of self-assembly protocols for the 4-patch disc of \f{fig_poly} using the microscopic network (left two columns) or the time network (right two columns). Networks instructed to maximize the number of 6-gons (columns 1 and 3) or 4-gons (columns 2 and 4) learn to promote the assembly of the 3.6.3.6 tiling or the rhombic structure. As in the other cases studied, the microscopic network is more effective than the time network. We show the top 5 protocols per generation, with the best shown in blue (dark gray).}
\end{figure*}

\begin{figure*}[] 
   \centering
\includegraphics[width=\linewidth]{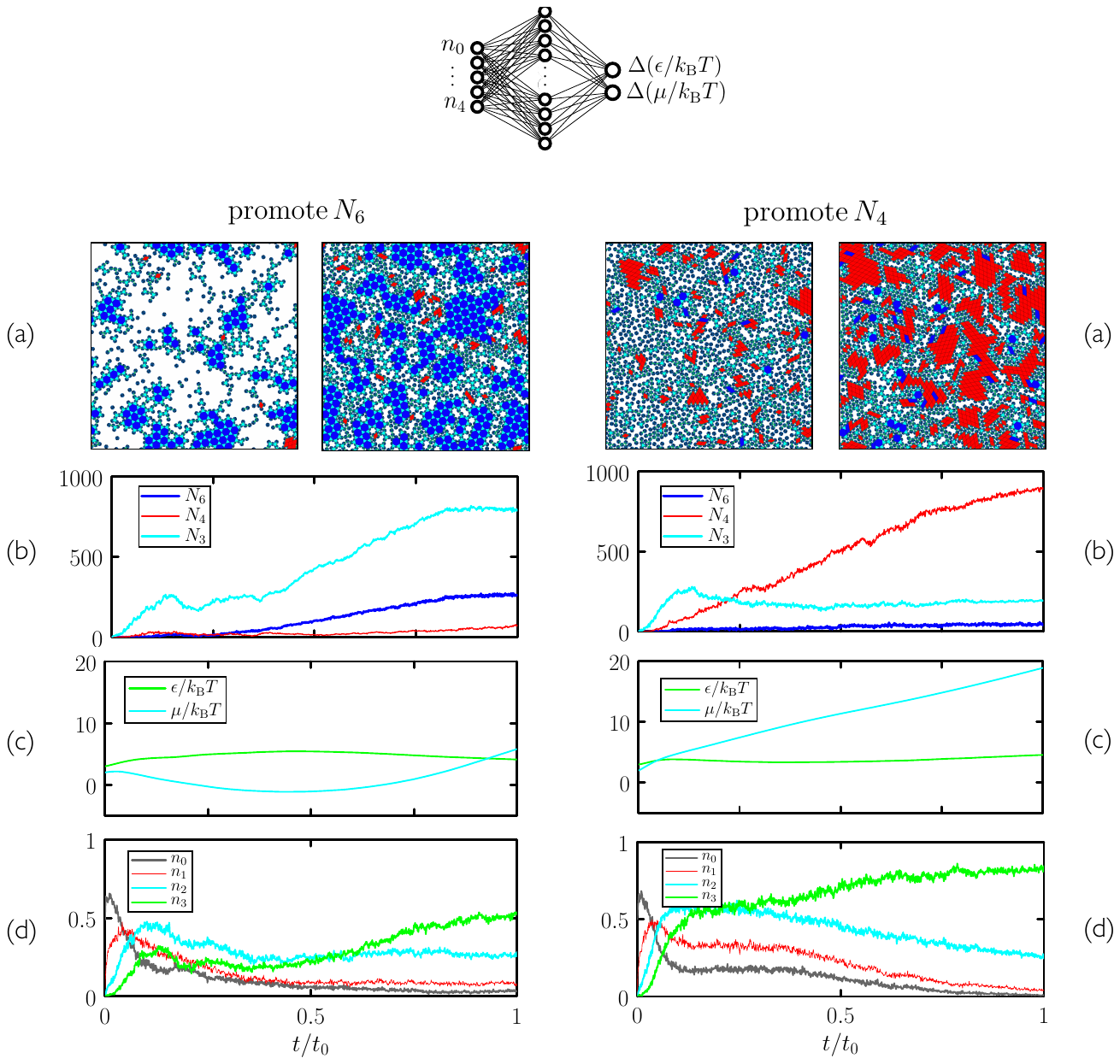} 
   \caption{\label{fig4} Generation-10 trajectories of the microscopic network from \f{fig5}, evolved to promote either 6-gons (left column) or 4-gons (right column). We show (a) time-ordered snapshots, (b) polygon counts, (c) neural network outputs, and (d) neural network inputs. In snapshots, 6-gons are dark blue (dark gray), 3-gons are light blue (light gray), and 4-gons are red (lighter gray). More detail can be seen in snapshots by enlarging them on a computer screen.}
\end{figure*}

\begin{figure}[] 
   \centering
\includegraphics[width=\linewidth]{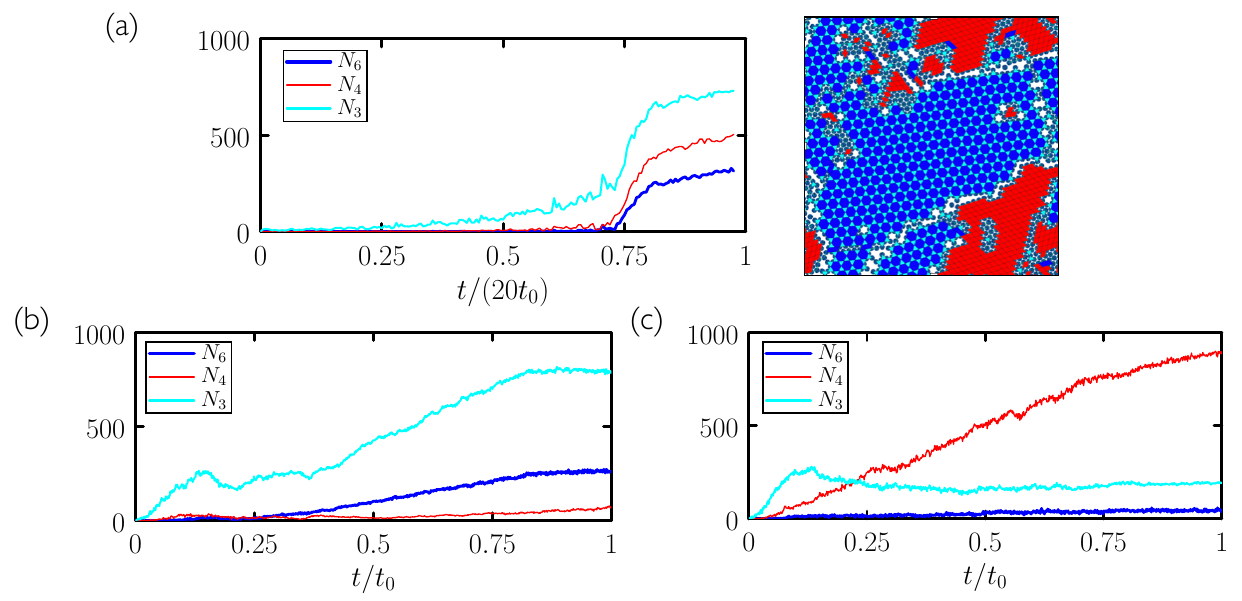} 
   \caption{\label{fig_mech} (a) Slow cooling of the 4-patch disc of \f{fig_poly} results in nucleation and growth of either polymorph type, in unpredictable quantities; see the gray circles in \f{fig_poly}(c). Here, both polymorphs appear. In the snapshot, 6-gons are dark blue (dark gray), 3-gons are light blue (light gray), and 4-gons are red (lighter gray). Note that the unit cells of these polymorphs are smaller and more mechanically rigid than that of the 3.12.12 structure of \f{fig_cool}, and so the 3.6.3.6 and rhombic polymorphs assemble better under slow cooling than the 3.12.12 structure. However, there exists no mechanism during slow nucleation and growth to reliably select one polymorph over the other. (b,c) By contrast, the non-nucleation mechanisms generated by neural networks evolved to promote 6-gons (b) or 4-gons (c) result in more predictable outcomes; see the blue (dark gray) and red (light gray) symbols in \f{fig_poly}(c).}
\end{figure}

Controlling the polymorph into which a set of molecules will self-assemble is a key consideration in industrial procedures such as drug crystallization\c{chen2011pharmaceutical,threlfall2000crystallisation,rodriguez1999significance,shekunov2000crystallization}. Here we show that evolutionary search can be used to find protocols able to direct the self-assembly of a set of model molecules into either of two competing polymorphs. In doing so, the procedure learns strategies that provide physical insight into the system under study.

In \f{fig_poly} we consider a 4-patch disc with angles $\pi/3$ and $2 \pi/3$ between patch bisectors. This disc can form a structure equivalent to the 3.6.3.6 Archimedean tiling (a tiling with two 3-gons and two 6-gons around each vertex), or a rhombic structure. Particles have equal energy within the bulk of each structure, and at zero pressure (the conditions experienced by a cluster growing in isolation in a substrate) there is no thermodynamic preference for one structure over the other. Independent trajectories generated under slow cooling (gray circles) therefore display nucleation of either or both polymorphs, in an unpredictable way (see also \f{fig_mech}) The 3.6.3.6 polymorph can be selected by making the patches chemically selective\c{whitelam2016minimal}, but here we do not do this. Instead, we show that evolutionary search can be used to develop protocols able to choose between these two polymorphs.

In \f{fig5} we consider evolutionary learning of self-assembly protocols using time- and microscopic networks instructed to promote either the parameter $N_4$ or the parameter $N_6$. In both cases we see steadily increasing counts, with generation, of the required order parameter, with the microscopic network again performing better. The lower two rows show the evolution of the strategies chosen by each network, with time- and microscopic networks learning qualitatively similar protocols for promotion of a given order parameter.

In \f{fig4} we show in more detail one trajectory per strategy obtained using generation-10 microscopic neural networks. Left-hand panels pertain to a trajectory produced by a neural network evolved to promote 6-gons, while right-hand panels pertain to a trajectory produced by a neural network evolved to promote 4-gons. In each case, neural networks have learned to promote one polymorph and so suppress the other. Both examples of assembly display defects and grain boundaries, but the specified polymorphs cover substantial parts of the substrate. In the case considered in \s{sec_ass} we already knew how to promote assembly, by cooling -- although the evolutionary protocol learned to do it more quickly and with higher fidelity -- but here we did not possess advance knowledge of how to do polymorph selection using protocol choice. 

In \f{fig4}, inspection of the snapshots (a), the polygon counts (b), and the control-parameter histories (c) provide insight into the selection strategies adopted by the networks. To select the 3.6.3.6 tiling (left panels) the network has induced a tendency for particles to leave the surface (small $\mu/\kt$) and for bonds to be moderately strong (moderate $\epsilon/\kt$). The balance of these things appears to be such that trimers (3-gons), in which each particle has two engaged bonds, can form. Trimers serve as a building block for the 3.6.3.6 structure, which then forms hierarchically as the chemical potential is increased (and the bond strength slightly {\em decreased}). By contrast, the rhombic structure appears to be unable to grow because it cannot form hierarchically from collections of rhombi (which also contain particles with two engaged bonds): growing beyond a single rhombus involves the addition of particles via only one engaged bond, and these particles are unstable, at early times, to dissociation. 

To select the rhombic structure (right panels) the network selects moderate bond strength and concentrates the substrate by driving $\mu/\kt$ large. In a dense environment it appears that the rhombic structure is more accessible kinetically than the more open 3.6.3.6 network. In addition, in a dense environment there is a thermodynamic preference for the more compact rhombic polymorph, a factor that may also contribute to selection of the latter. Note that simply causing $\mu/\kt$ to increase with time is not sufficient to produce the rhombic polymorph in high yield: early-generation networks adopt just such a strategy, but high yield for later generations requires a particular balance of bond strength and chemical potential.

The microscopic network receives information periodically from the system, but the information it receives -- the number of particles with certain numbers of engaged bonds -- does not distinguish between the bulk forms of the two polymorphs. Networks must therefore learn the relationships between these inputs, their resulting actions, and the final-time order parameter. The time network learns qualitatively similar protocols, albeit with slightly less effectiveness, with no access to the microscopic state of the system.

Returning to \f{fig_poly}(c), we show the results of 50 independent trajectories of length $t_0$ carried out using a single generation-10 microscopic network evolved to promote 6-gons (blue hexagons), and the results of 50 independent trajectories of length $t_0$ carried out using a single generation-10 microscopic network evolved to promote 4-gons. In both cases the networks reliably promote one polymorph and suppress the other, in contrast to slow-cooling simulations whose outcome is unpredictable. In this case the conventional nucleation-and-growth pathway induced by slow cooling provides no control over polymorph selection, while the pathways induced by the neural networks -- one of which is strongly hierarchical -- do. In addition, as in \s{sec_ass}, assembly under the network protocols is much faster than under slow cooling: see \f{fig_mech}.

\section{Conclusions}
\label{sec_conc}

We have shown that a self-assembly kinetic yield net trained by evolutionary reinforcement learning\c{GA,GA2,lehman2018more,salimans2017evolution,zhang2017relationship,lehman2018safe,conti2018improving,Guber} can control self-assembly protocols in molecular simulations. Networks learn to promote the assembly of desired structures, or choose between polymorphs. In the first case, networks reproduce the structures produced by previously-known protocols, but faster and with higher fidelity; in the second case they identify strategies previously unknown, and from which we can extract physical insight. Networks that take as input only the elapsed time of the protocol are effective, and networks that take as input microscopic information from the system are more so. This comparison indicates that this scheme can be applied to a wide range of experiments, regardless of how much microscopic information is available as assembly proceeds.  

The problem we have addressed falls in the category of reinforcement learning in the sense that the neural network learns to perform actions (choosing new values of the control parameters) given an observation. The evolutionary approach we have applied to this problem requires only the assessment of a desired order parameter (here the polygon count $N_\alpha$) at the end of a trajectory. This is an important feature because in self-assembly the best-performing trajectories at short times are not necessarily the best-performing trajectories at the desired observation time: see e.g. \f{fig2_supp}. Self-assembly is inherently a ``sparse-reward'' problem. For this reason it is not obvious that value-based reinforcement-learning methods\c{sutton2018reinforcement} are ideally suited to a problem such as self-assembly: rewarding ``good'' configurations at early times may not result in favorable outcomes at later times. This is only speculation on our part, however; which of the many ways of doing reinforcement learning is best for self-assembly is an open question.

Our results demonstrate proof of principle, and can be extended or adapted in several ways. We allow networks to act 1000 times per trajectory, in order to mimic a system in which we have only occasional control; the influence of a network could be increased by allowing it to act more frequently. We have chosen the hyperparameters of our scheme (mutation step size, neural network width, network activation functions, number of trajectories per generation) using values that seemed reasonable and that we subsequently observed to work, but these could be optimized (potentially by evolutionary search). 

We end by noting that the scheme we have used is simple to implement. The network architectures we have used are standard and can be straightforwardly adapted to handle an arbitrary number of inputs (system data) and outputs (changes of system control parameters). The mutation protocol is simple to implement. In addition, we have shown that learning can be effective using a modest number of trajectories (50) per generation. The evolutionary scheme should therefore be applicable to a broad range of experimental or computational systems. The results shown here have been achieved with no human input beyond the specification of which order parameter to promote, pointing the way to the design of synthesis protocols by artificial intelligence.

{\em Acknowledgments} -- This work was performed as part of a user project at the Molecular Foundry, Lawrence Berkeley National Laboratory, supported by the Office of Science, Office of Basic Energy Sciences, of the U.S. Department of Energy under Contract No. DE-AC02--05CH11231. I.T. performed work at the National Research Council of Canada under the auspices of the AI4D Program. 

%\bibliography{bib}

\begin{thebibliography}{74}%
\makeatletter
\providecommand \@ifxundefined [1]{%
 \@ifx{#1\undefined}
}%
\providecommand \@ifnum [1]{%
 \ifnum #1\expandafter \@firstoftwo
 \else \expandafter \@secondoftwo
 \fi
}%
\providecommand \@ifx [1]{%
 \ifx #1\expandafter \@firstoftwo
 \else \expandafter \@secondoftwo
 \fi
}%
\providecommand \natexlab [1]{#1}%
\providecommand \enquote  [1]{``#1''}%
\providecommand \bibnamefont  [1]{#1}%
\providecommand \bibfnamefont [1]{#1}%
\providecommand \citenamefont [1]{#1}%
\providecommand \href@noop [0]{\@secondoftwo}%
\providecommand \href [0]{\begingroup \@sanitize@url \@href}%
\providecommand \@href[1]{\@@startlink{#1}\@@href}%
\providecommand \@@href[1]{\endgroup#1\@@endlink}%
\providecommand \@sanitize@url [0]{\catcode `\\12\catcode `\$12\catcode
  `\&12\catcode `\#12\catcode `\^12\catcode `\_12\catcode `\%12\relax}%
\providecommand \@@startlink[1]{}%
\providecommand \@@endlink[0]{}%
\providecommand \url  [0]{\begingroup\@sanitize@url \@url }%
\providecommand \@url [1]{\endgroup\@href {#1}{\urlprefix }}%
\providecommand \urlprefix  [0]{URL }%
\providecommand \Eprint [0]{\href }%
\providecommand \doibase [0]{http://dx.doi.org/}%
\providecommand \selectlanguage [0]{\@gobble}%
\providecommand \bibinfo  [0]{\@secondoftwo}%
\providecommand \bibfield  [0]{\@secondoftwo}%
\providecommand \translation [1]{[#1]}%
\providecommand \BibitemOpen [0]{}%
\providecommand \bibitemStop [0]{}%
\providecommand \bibitemNoStop [0]{.\EOS\space}%
\providecommand \EOS [0]{\spacefactor3000\relax}%
\providecommand \BibitemShut  [1]{\csname bibitem#1\endcsname}%
\let\auto@bib@innerbib\@empty
%</preamble>
\bibitem [{\citenamefont {Whitesides}\ \emph {et~al.}(1991)\citenamefont
  {Whitesides}, \citenamefont {Mathias},\ and\ \citenamefont
  {Seto}}]{whitesides1991molecular}%
  \BibitemOpen
  \bibfield  {author} {\bibinfo {author} {\bibfnamefont {George~M}\
  \bibnamefont {Whitesides}}, \bibinfo {author} {\bibfnamefont {John~P}\
  \bibnamefont {Mathias}}, \ and\ \bibinfo {author} {\bibfnamefont
  {Christopher~T}\ \bibnamefont {Seto}},\ }\bibfield  {title} {\enquote
  {\bibinfo {title} {Molecular self-assembly and nanochemistry: A chemical
  strategy for the synthesis of nanostructures},}\ }\href@noop {} {\bibfield
  {journal} {\bibinfo  {journal} {Science}\ }\textbf {\bibinfo {volume}
  {254}},\ \bibinfo {pages} {1312--1319} (\bibinfo {year} {1991})}\BibitemShut
  {NoStop}%
\bibitem [{\citenamefont {Biancaniello}\ \emph {et~al.}(2005)\citenamefont
  {Biancaniello}, \citenamefont {Kim},\ and\ \citenamefont
  {Crocker}}]{biancaniello2005colloidal}%
  \BibitemOpen
  \bibfield  {author} {\bibinfo {author} {\bibfnamefont {Paul~L}\ \bibnamefont
  {Biancaniello}}, \bibinfo {author} {\bibfnamefont {Anthony~J}\ \bibnamefont
  {Kim}}, \ and\ \bibinfo {author} {\bibfnamefont {John~C}\ \bibnamefont
  {Crocker}},\ }\bibfield  {title} {\enquote {\bibinfo {title} {Colloidal
  interactions and self-assembly using dna hybridization},}\ }\href@noop {}
  {\bibfield  {journal} {\bibinfo  {journal} {Physical Review Letters}\
  }\textbf {\bibinfo {volume} {94}},\ \bibinfo {pages} {058302} (\bibinfo
  {year} {2005})}\BibitemShut {NoStop}%
\bibitem [{\citenamefont {Park}\ \emph {et~al.}(2008)\citenamefont {Park},
  \citenamefont {Lytton-Jean}, \citenamefont {Lee}, \citenamefont {Weigand},
  \citenamefont {Schatz},\ and\ \citenamefont {Mirkin}}]{park2008dna}%
  \BibitemOpen
  \bibfield  {author} {\bibinfo {author} {\bibfnamefont {Sung~Yong}\
  \bibnamefont {Park}}, \bibinfo {author} {\bibfnamefont {Abigail~KR}\
  \bibnamefont {Lytton-Jean}}, \bibinfo {author} {\bibfnamefont {Byeongdu}\
  \bibnamefont {Lee}}, \bibinfo {author} {\bibfnamefont {Steven}\ \bibnamefont
  {Weigand}}, \bibinfo {author} {\bibfnamefont {George~C}\ \bibnamefont
  {Schatz}}, \ and\ \bibinfo {author} {\bibfnamefont {Chad~A}\ \bibnamefont
  {Mirkin}},\ }\bibfield  {title} {\enquote {\bibinfo {title} {Dna-programmable
  nanoparticle crystallization},}\ }\href@noop {} {\bibfield  {journal}
  {\bibinfo  {journal} {Nature}\ }\textbf {\bibinfo {volume} {451}},\ \bibinfo
  {pages} {553--556} (\bibinfo {year} {2008})}\BibitemShut {NoStop}%
\bibitem [{\citenamefont {Nykypanchuk}\ \emph {et~al.}(2008)\citenamefont
  {Nykypanchuk}, \citenamefont {Maye}, \citenamefont {van~der Lelie},\ and\
  \citenamefont {Gang}}]{nykypanchuk2008dna}%
  \BibitemOpen
  \bibfield  {author} {\bibinfo {author} {\bibfnamefont {Dmytro}\ \bibnamefont
  {Nykypanchuk}}, \bibinfo {author} {\bibfnamefont {Mathew~M}\ \bibnamefont
  {Maye}}, \bibinfo {author} {\bibfnamefont {Daniel}\ \bibnamefont {van~der
  Lelie}}, \ and\ \bibinfo {author} {\bibfnamefont {Oleg}\ \bibnamefont
  {Gang}},\ }\bibfield  {title} {\enquote {\bibinfo {title} {Dna-guided
  crystallization of colloidal nanoparticles},}\ }\href@noop {} {\bibfield
  {journal} {\bibinfo  {journal} {Nature}\ }\textbf {\bibinfo {volume} {451}},\
  \bibinfo {pages} {549--552} (\bibinfo {year} {2008})}\BibitemShut {NoStop}%
\bibitem [{\citenamefont {Ke}\ \emph {et~al.}(2012)\citenamefont {Ke},
  \citenamefont {Ong}, \citenamefont {Shih},\ and\ \citenamefont
  {Yin}}]{ke2012three}%
  \BibitemOpen
  \bibfield  {author} {\bibinfo {author} {\bibfnamefont {Yonggang}\
  \bibnamefont {Ke}}, \bibinfo {author} {\bibfnamefont {Luvena~L}\ \bibnamefont
  {Ong}}, \bibinfo {author} {\bibfnamefont {William~M}\ \bibnamefont {Shih}}, \
  and\ \bibinfo {author} {\bibfnamefont {Peng}\ \bibnamefont {Yin}},\
  }\bibfield  {title} {\enquote {\bibinfo {title} {Three-dimensional structures
  self-assembled from dna bricks},}\ }\href@noop {} {\bibfield  {journal}
  {\bibinfo  {journal} {Science}\ }\textbf {\bibinfo {volume} {338}},\ \bibinfo
  {pages} {1177--1183} (\bibinfo {year} {2012})}\BibitemShut {NoStop}%
\bibitem [{\citenamefont {Pfeifer}\ and\ \citenamefont
  {Sacc{\`a}}(2018)}]{pfeifer2018synthetic}%
  \BibitemOpen
  \bibfield  {author} {\bibinfo {author} {\bibfnamefont {Wolfgang}\
  \bibnamefont {Pfeifer}}\ and\ \bibinfo {author} {\bibfnamefont {Barbara}\
  \bibnamefont {Sacc{\`a}}},\ }\bibfield  {title} {\enquote {\bibinfo {title}
  {Synthetic dna filaments: from design to applications},}\ }\href@noop {}
  {\bibfield  {journal} {\bibinfo  {journal} {Biological Chemistry}\ }\textbf
  {\bibinfo {volume} {399}},\ \bibinfo {pages} {773--785} (\bibinfo {year}
  {2018})}\BibitemShut {NoStop}%
\bibitem [{\citenamefont {Doye}\ \emph {et~al.}(2004)\citenamefont {Doye},
  \citenamefont {Louis},\ and\ \citenamefont
  {Vendruscolo}}]{doye2004inhibition}%
  \BibitemOpen
  \bibfield  {author} {\bibinfo {author} {\bibfnamefont {J.~P.~K.}\
  \bibnamefont {Doye}}, \bibinfo {author} {\bibfnamefont {A.~A.}\ \bibnamefont
  {Louis}}, \ and\ \bibinfo {author} {\bibfnamefont {M.}~\bibnamefont
  {Vendruscolo}},\ }\bibfield  {title} {\enquote {\bibinfo {title} {Inhibition
  of protein crystallization by evolutionary negative design},}\ }\href@noop {}
  {\bibfield  {journal} {\bibinfo  {journal} {Physical Biology}\ }\textbf
  {\bibinfo {volume} {1}},\ \bibinfo {pages} {P9} (\bibinfo {year}
  {2004})}\BibitemShut {NoStop}%
\bibitem [{\citenamefont {Romano}\ and\ \citenamefont
  {Sciortino}(2011)}]{romano2011colloidal}%
  \BibitemOpen
  \bibfield  {author} {\bibinfo {author} {\bibfnamefont {Flavio}\ \bibnamefont
  {Romano}}\ and\ \bibinfo {author} {\bibfnamefont {Francesco}\ \bibnamefont
  {Sciortino}},\ }\bibfield  {title} {\enquote {\bibinfo {title} {Colloidal
  self-assembly: patchy from the bottom up},}\ }\href@noop {} {\bibfield
  {journal} {\bibinfo  {journal} {Nature materials}\ }\textbf {\bibinfo
  {volume} {10}},\ \bibinfo {pages} {171} (\bibinfo {year} {2011})}\BibitemShut
  {NoStop}%
\bibitem [{\citenamefont {Glotzer}\ \emph {et~al.}(2004)\citenamefont
  {Glotzer}, \citenamefont {Solomon},\ and\ \citenamefont
  {Kotov}}]{glotzer2004self}%
  \BibitemOpen
  \bibfield  {author} {\bibinfo {author} {\bibfnamefont {SC}~\bibnamefont
  {Glotzer}}, \bibinfo {author} {\bibfnamefont {MJ}~\bibnamefont {Solomon}}, \
  and\ \bibinfo {author} {\bibfnamefont {Nicholas~A}\ \bibnamefont {Kotov}},\
  }\bibfield  {title} {\enquote {\bibinfo {title} {Self-assembly: From
  nanoscale to microscale colloids},}\ }\href@noop {} {\bibfield  {journal}
  {\bibinfo  {journal} {AIChE Journal}\ }\textbf {\bibinfo {volume} {50}},\
  \bibinfo {pages} {2978--2985} (\bibinfo {year} {2004})}\BibitemShut {NoStop}%
\bibitem [{\citenamefont {Doye}\ \emph
  {et~al.}(2007{\natexlab{a}})\citenamefont {Doye}, \citenamefont {Louis},
  \citenamefont {Lin}, \citenamefont {Allen}, \citenamefont {Noya},
  \citenamefont {Wilber}, \citenamefont {Kok},\ and\ \citenamefont
  {Lyus}}]{doye2007condensed}%
  \BibitemOpen
  \bibfield  {author} {\bibinfo {author} {\bibfnamefont {J.~P.~K.}\
  \bibnamefont {Doye}}, \bibinfo {author} {\bibfnamefont {A.~A.}\ \bibnamefont
  {Louis}}, \bibinfo {author} {\bibfnamefont {I.~C.}\ \bibnamefont {Lin}},
  \bibinfo {author} {\bibfnamefont {L.~R.}\ \bibnamefont {Allen}}, \bibinfo
  {author} {\bibfnamefont {E.~G.}\ \bibnamefont {Noya}}, \bibinfo {author}
  {\bibfnamefont {A.~W.}\ \bibnamefont {Wilber}}, \bibinfo {author}
  {\bibfnamefont {H.~C.}\ \bibnamefont {Kok}}, \ and\ \bibinfo {author}
  {\bibfnamefont {R.}~\bibnamefont {Lyus}},\ }\bibfield  {title} {\enquote
  {\bibinfo {title} {{Controlling crystallization and its absence: Proteins,
  colloids and patchy models}},}\ }\href@noop {} {\bibfield  {journal}
  {\bibinfo  {journal} {Physical Chemistry Chemical Physics}\ }\textbf
  {\bibinfo {volume} {9}},\ \bibinfo {pages} {2197--2205} (\bibinfo {year}
  {2007}{\natexlab{a}})}\BibitemShut {NoStop}%
\bibitem [{\citenamefont {Rapaport}(2010)}]{rapaport2010modeling}%
  \BibitemOpen
  \bibfield  {author} {\bibinfo {author} {\bibfnamefont {D.~C.}\ \bibnamefont
  {Rapaport}},\ }\bibfield  {title} {\enquote {\bibinfo {title} {Modeling
  capsid self-assembly: design and analysis},}\ }\href@noop {} {\bibfield
  {journal} {\bibinfo  {journal} {Phys. Biol.}\ }\textbf {\bibinfo {volume}
  {7}},\ \bibinfo {pages} {045001} (\bibinfo {year} {2010})}\BibitemShut
  {NoStop}%
\bibitem [{\citenamefont {Reinhardt}\ and\ \citenamefont
  {Frenkel}(2014)}]{reinhardt2014numerical}%
  \BibitemOpen
  \bibfield  {author} {\bibinfo {author} {\bibfnamefont {Aleks}\ \bibnamefont
  {Reinhardt}}\ and\ \bibinfo {author} {\bibfnamefont {Daan}\ \bibnamefont
  {Frenkel}},\ }\bibfield  {title} {\enquote {\bibinfo {title} {Numerical
  evidence for nucleated self-assembly of dna brick structures},}\ }\href@noop
  {} {\bibfield  {journal} {\bibinfo  {journal} {Physical Review Letters}\
  }\textbf {\bibinfo {volume} {112}},\ \bibinfo {pages} {238103} (\bibinfo
  {year} {2014})}\BibitemShut {NoStop}%
\bibitem [{\citenamefont {De~Yoreo}\ \emph {et~al.}(2015)\citenamefont
  {De~Yoreo}, \citenamefont {Gilbert}, \citenamefont {Sommerdijk},
  \citenamefont {Penn}, \citenamefont {Whitelam}, \citenamefont {Joester},
  \citenamefont {Zhang}, \citenamefont {Rimer}, \citenamefont {Navrotsky},
  \citenamefont {Banfield} \emph {et~al.}}]{de2015crystallization}%
  \BibitemOpen
  \bibfield  {author} {\bibinfo {author} {\bibfnamefont {James~J}\ \bibnamefont
  {De~Yoreo}}, \bibinfo {author} {\bibfnamefont {Pupa~UPA}\ \bibnamefont
  {Gilbert}}, \bibinfo {author} {\bibfnamefont {Nico~AJM}\ \bibnamefont
  {Sommerdijk}}, \bibinfo {author} {\bibfnamefont {R~Lee}\ \bibnamefont
  {Penn}}, \bibinfo {author} {\bibfnamefont {Stephen}\ \bibnamefont
  {Whitelam}}, \bibinfo {author} {\bibfnamefont {Derk}\ \bibnamefont
  {Joester}}, \bibinfo {author} {\bibfnamefont {Hengzhong}\ \bibnamefont
  {Zhang}}, \bibinfo {author} {\bibfnamefont {Jeffrey~D}\ \bibnamefont
  {Rimer}}, \bibinfo {author} {\bibfnamefont {Alexandra}\ \bibnamefont
  {Navrotsky}}, \bibinfo {author} {\bibfnamefont {Jillian~F}\ \bibnamefont
  {Banfield}},  \emph {et~al.},\ }\bibfield  {title} {\enquote {\bibinfo
  {title} {Crystallization by particle attachment in synthetic, biogenic, and
  geologic environments},}\ }\href@noop {} {\bibfield  {journal} {\bibinfo
  {journal} {Science}\ }\textbf {\bibinfo {volume} {349}},\ \bibinfo {pages}
  {aaa6760} (\bibinfo {year} {2015})}\BibitemShut {NoStop}%
\bibitem [{\citenamefont {Murugan}\ \emph {et~al.}(2015)\citenamefont
  {Murugan}, \citenamefont {Zou},\ and\ \citenamefont
  {Brenner}}]{murugan2015undesired}%
  \BibitemOpen
  \bibfield  {author} {\bibinfo {author} {\bibfnamefont {Arvind}\ \bibnamefont
  {Murugan}}, \bibinfo {author} {\bibfnamefont {James}\ \bibnamefont {Zou}}, \
  and\ \bibinfo {author} {\bibfnamefont {Michael~P}\ \bibnamefont {Brenner}},\
  }\bibfield  {title} {\enquote {\bibinfo {title} {Undesired usage and the
  robust self-assembly of heterogeneous structures},}\ }\href@noop {}
  {\bibfield  {journal} {\bibinfo  {journal} {Nature Communications}\ }\textbf
  {\bibinfo {volume} {6}} (\bibinfo {year} {2015})}\BibitemShut {NoStop}%
\bibitem [{\citenamefont {Whitelam}\ and\ \citenamefont
  {Jack}(2015)}]{whitelam2015statistical}%
  \BibitemOpen
  \bibfield  {author} {\bibinfo {author} {\bibfnamefont {Stephen}\ \bibnamefont
  {Whitelam}}\ and\ \bibinfo {author} {\bibfnamefont {Robert~L}\ \bibnamefont
  {Jack}},\ }\bibfield  {title} {\enquote {\bibinfo {title} {The statistical
  mechanics of dynamic pathways to self-assembly},}\ }\href@noop {} {\bibfield
  {journal} {\bibinfo  {journal} {Annual Review of Physical Chemistry}\
  }\textbf {\bibinfo {volume} {66}},\ \bibinfo {pages} {143--163} (\bibinfo
  {year} {2015})}\BibitemShut {NoStop}%
\bibitem [{\citenamefont {Nguyen}\ and\ \citenamefont
  {Vaikuntanathan}(2016)}]{nguyen2016design}%
  \BibitemOpen
  \bibfield  {author} {\bibinfo {author} {\bibfnamefont {Michael}\ \bibnamefont
  {Nguyen}}\ and\ \bibinfo {author} {\bibfnamefont {Suriyanarayanan}\
  \bibnamefont {Vaikuntanathan}},\ }\bibfield  {title} {\enquote {\bibinfo
  {title} {Design principles for nonequilibrium self-assembly},}\ }\href@noop
  {} {\bibfield  {journal} {\bibinfo  {journal} {Proceedings of the National
  Academy of Sciences}\ }\textbf {\bibinfo {volume} {113}},\ \bibinfo {pages}
  {14231--14236} (\bibinfo {year} {2016})}\BibitemShut {NoStop}%
\bibitem [{\citenamefont {Whitelam}\ \emph
  {et~al.}(2014{\natexlab{a}})\citenamefont {Whitelam}, \citenamefont
  {Hedges},\ and\ \citenamefont {Schmit}}]{whitelam2014self}%
  \BibitemOpen
  \bibfield  {author} {\bibinfo {author} {\bibfnamefont {Stephen}\ \bibnamefont
  {Whitelam}}, \bibinfo {author} {\bibfnamefont {Lester~O}\ \bibnamefont
  {Hedges}}, \ and\ \bibinfo {author} {\bibfnamefont {Jeremy~D}\ \bibnamefont
  {Schmit}},\ }\bibfield  {title} {\enquote {\bibinfo {title} {Self-assembly at
  a nonequilibrium critical point},}\ }\href@noop {} {\bibfield  {journal}
  {\bibinfo  {journal} {Physical Review Letters}\ }\textbf {\bibinfo {volume}
  {112}},\ \bibinfo {pages} {155504} (\bibinfo {year}
  {2014}{\natexlab{a}})}\BibitemShut {NoStop}%
\bibitem [{\citenamefont {Jadrich}\ \emph {et~al.}(2017)\citenamefont
  {Jadrich}, \citenamefont {Lindquist},\ and\ \citenamefont
  {Truskett}}]{jadrich2017probabilistic}%
  \BibitemOpen
  \bibfield  {author} {\bibinfo {author} {\bibfnamefont {RB}~\bibnamefont
  {Jadrich}}, \bibinfo {author} {\bibfnamefont {BA}~\bibnamefont {Lindquist}},
  \ and\ \bibinfo {author} {\bibfnamefont {TM}~\bibnamefont {Truskett}},\
  }\bibfield  {title} {\enquote {\bibinfo {title} {Probabilistic inverse design
  for self-assembling materials},}\ }\href@noop {} {\bibfield  {journal}
  {\bibinfo  {journal} {The Journal of Chemical Physics}\ }\textbf {\bibinfo
  {volume} {146}},\ \bibinfo {pages} {184103} (\bibinfo {year}
  {2017})}\BibitemShut {NoStop}%
\bibitem [{\citenamefont {Lutsko}(2019)}]{lutsko2019crystals}%
  \BibitemOpen
  \bibfield  {author} {\bibinfo {author} {\bibfnamefont {James~F}\ \bibnamefont
  {Lutsko}},\ }\bibfield  {title} {\enquote {\bibinfo {title} {How crystals
  form: A theory of nucleation pathways},}\ }\href@noop {} {\bibfield
  {journal} {\bibinfo  {journal} {Science advances}\ }\textbf {\bibinfo
  {volume} {5}},\ \bibinfo {pages} {eaav7399} (\bibinfo {year}
  {2019})}\BibitemShut {NoStop}%
\bibitem [{\citenamefont {Fan}\ and\ \citenamefont
  {Grunwald}(2019)}]{fan2019orientational}%
  \BibitemOpen
  \bibfield  {author} {\bibinfo {author} {\bibfnamefont {Zhaochuan}\
  \bibnamefont {Fan}}\ and\ \bibinfo {author} {\bibfnamefont {Michael}\
  \bibnamefont {Grunwald}},\ }\bibfield  {title} {\enquote {\bibinfo {title}
  {Orientational order in self-assembled nanocrystal superlattices},}\
  }\href@noop {} {\bibfield  {journal} {\bibinfo  {journal} {Journal of the
  American Chemical Society}\ }\textbf {\bibinfo {volume} {141}},\ \bibinfo
  {pages} {1980--1988} (\bibinfo {year} {2019})}\BibitemShut {NoStop}%
\bibitem [{\citenamefont {Chen}\ \emph {et~al.}(2011)\citenamefont {Chen},
  \citenamefont {Sarma}, \citenamefont {Evans},\ and\ \citenamefont
  {Myerson}}]{chen2011pharmaceutical}%
  \BibitemOpen
  \bibfield  {author} {\bibinfo {author} {\bibfnamefont {Jie}\ \bibnamefont
  {Chen}}, \bibinfo {author} {\bibfnamefont {Bipul}\ \bibnamefont {Sarma}},
  \bibinfo {author} {\bibfnamefont {James~MB}\ \bibnamefont {Evans}}, \ and\
  \bibinfo {author} {\bibfnamefont {Allan~S}\ \bibnamefont {Myerson}},\
  }\bibfield  {title} {\enquote {\bibinfo {title} {Pharmaceutical
  crystallization},}\ }\href@noop {} {\bibfield  {journal} {\bibinfo  {journal}
  {Crystal growth \& design}\ }\textbf {\bibinfo {volume} {11}},\ \bibinfo
  {pages} {887--895} (\bibinfo {year} {2011})}\BibitemShut {NoStop}%
\bibitem [{\citenamefont {Threlfall}(2000)}]{threlfall2000crystallisation}%
  \BibitemOpen
  \bibfield  {author} {\bibinfo {author} {\bibfnamefont {Terry}\ \bibnamefont
  {Threlfall}},\ }\bibfield  {title} {\enquote {\bibinfo {title}
  {Crystallisation of polymorphs: thermodynamic insight into the role of
  solvent},}\ }\href@noop {} {\bibfield  {journal} {\bibinfo  {journal}
  {Organic Process Research \& Development}\ }\textbf {\bibinfo {volume} {4}},\
  \bibinfo {pages} {384--390} (\bibinfo {year} {2000})}\BibitemShut {NoStop}%
\bibitem [{\citenamefont {Rodr{\'\i}guez-hornedo}\ and\ \citenamefont
  {Murphy}(1999)}]{rodriguez1999significance}%
  \BibitemOpen
  \bibfield  {author} {\bibinfo {author} {\bibfnamefont {Na{\'\i}r}\
  \bibnamefont {Rodr{\'\i}guez-hornedo}}\ and\ \bibinfo {author} {\bibfnamefont
  {Denette}\ \bibnamefont {Murphy}},\ }\bibfield  {title} {\enquote {\bibinfo
  {title} {Significance of controlling crystallization mechanisms and kinetics
  in pharmaceutical systems},}\ }\href@noop {} {\bibfield  {journal} {\bibinfo
  {journal} {Journal of pharmaceutical sciences}\ }\textbf {\bibinfo {volume}
  {88}},\ \bibinfo {pages} {651--660} (\bibinfo {year} {1999})}\BibitemShut
  {NoStop}%
\bibitem [{\citenamefont {Shekunov}\ and\ \citenamefont
  {York}(2000)}]{shekunov2000crystallization}%
  \BibitemOpen
  \bibfield  {author} {\bibinfo {author} {\bibfnamefont {B~Yu}\ \bibnamefont
  {Shekunov}}\ and\ \bibinfo {author} {\bibfnamefont {P}~\bibnamefont {York}},\
  }\bibfield  {title} {\enquote {\bibinfo {title} {Crystallization processes in
  pharmaceutical technology and drug delivery design},}\ }\href@noop {}
  {\bibfield  {journal} {\bibinfo  {journal} {Journal of crystal growth}\
  }\textbf {\bibinfo {volume} {211}},\ \bibinfo {pages} {122--136} (\bibinfo
  {year} {2000})}\BibitemShut {NoStop}%
\bibitem [{\citenamefont {Sutton}\ and\ \citenamefont
  {Barto}(2018)}]{sutton2018reinforcement}%
  \BibitemOpen
  \bibfield  {author} {\bibinfo {author} {\bibfnamefont {Richard~S}\
  \bibnamefont {Sutton}}\ and\ \bibinfo {author} {\bibfnamefont {Andrew~G}\
  \bibnamefont {Barto}},\ }\href@noop {} {\emph {\bibinfo {title}
  {Reinforcement learning: An introduction}}}\ (\bibinfo {year}
  {2018})\BibitemShut {NoStop}%
\bibitem [{\citenamefont {Watkins}\ and\ \citenamefont {Dayan}(1992)}]{QL}%
  \BibitemOpen
  \bibfield  {author} {\bibinfo {author} {\bibfnamefont {Christopher~JCH}\
  \bibnamefont {Watkins}}\ and\ \bibinfo {author} {\bibfnamefont {Peter}\
  \bibnamefont {Dayan}},\ }\bibfield  {title} {\enquote {\bibinfo {title}
  {Q-learning},}\ }\href@noop {} {\bibfield  {journal} {\bibinfo  {journal}
  {Machine learning}\ }\textbf {\bibinfo {volume} {8}},\ \bibinfo {pages}
  {279--292} (\bibinfo {year} {1992})}\BibitemShut {NoStop}%
\bibitem [{\citenamefont {Mnih}\ \emph {et~al.}(2013)\citenamefont {Mnih},
  \citenamefont {Kavukcuoglu}, \citenamefont {Silver}, \citenamefont {Graves},
  \citenamefont {Antonoglou}, \citenamefont {Wierstra},\ and\ \citenamefont
  {Riedmiller}}]{DQN}%
  \BibitemOpen
  \bibfield  {author} {\bibinfo {author} {\bibfnamefont {Volodymyr}\
  \bibnamefont {Mnih}}, \bibinfo {author} {\bibfnamefont {Koray}\ \bibnamefont
  {Kavukcuoglu}}, \bibinfo {author} {\bibfnamefont {David}\ \bibnamefont
  {Silver}}, \bibinfo {author} {\bibfnamefont {Alex}\ \bibnamefont {Graves}},
  \bibinfo {author} {\bibfnamefont {Ioannis}\ \bibnamefont {Antonoglou}},
  \bibinfo {author} {\bibfnamefont {Daan}\ \bibnamefont {Wierstra}}, \ and\
  \bibinfo {author} {\bibfnamefont {Martin}\ \bibnamefont {Riedmiller}},\
  }\bibfield  {title} {\enquote {\bibinfo {title} {Playing atari with deep
  reinforcement learning},}\ }\href@noop {} {\bibfield  {journal} {\bibinfo
  {journal} {arXiv preprint arXiv:1312.5602}\ } (\bibinfo {year}
  {2013})}\BibitemShut {NoStop}%
\bibitem [{\citenamefont {Mnih}\ \emph {et~al.}(2015)\citenamefont {Mnih},
  \citenamefont {Kavukcuoglu}, \citenamefont {Silver}, \citenamefont {Rusu},
  \citenamefont {Veness}, \citenamefont {Bellemare}, \citenamefont {Graves},
  \citenamefont {Riedmiller}, \citenamefont {Fidjeland}, \citenamefont
  {Ostrovski} \emph {et~al.}}]{Atari}%
  \BibitemOpen
  \bibfield  {author} {\bibinfo {author} {\bibfnamefont {Volodymyr}\
  \bibnamefont {Mnih}}, \bibinfo {author} {\bibfnamefont {Koray}\ \bibnamefont
  {Kavukcuoglu}}, \bibinfo {author} {\bibfnamefont {David}\ \bibnamefont
  {Silver}}, \bibinfo {author} {\bibfnamefont {Andrei~A}\ \bibnamefont {Rusu}},
  \bibinfo {author} {\bibfnamefont {Joel}\ \bibnamefont {Veness}}, \bibinfo
  {author} {\bibfnamefont {Marc~G}\ \bibnamefont {Bellemare}}, \bibinfo
  {author} {\bibfnamefont {Alex}\ \bibnamefont {Graves}}, \bibinfo {author}
  {\bibfnamefont {Martin}\ \bibnamefont {Riedmiller}}, \bibinfo {author}
  {\bibfnamefont {Andreas~K}\ \bibnamefont {Fidjeland}}, \bibinfo {author}
  {\bibfnamefont {Georg}\ \bibnamefont {Ostrovski}},  \emph {et~al.},\
  }\bibfield  {title} {\enquote {\bibinfo {title} {Human-level control through
  deep reinforcement learning},}\ }\href@noop {} {\bibfield  {journal}
  {\bibinfo  {journal} {Nature}\ }\textbf {\bibinfo {volume} {518}},\ \bibinfo
  {pages} {529} (\bibinfo {year} {2015})}\BibitemShut {NoStop}%
\bibitem [{\citenamefont {Bellemare}\ \emph {et~al.}(2013)\citenamefont
  {Bellemare}, \citenamefont {Naddaf}, \citenamefont {Veness},\ and\
  \citenamefont {Bowling}}]{Atari2600}%
  \BibitemOpen
  \bibfield  {author} {\bibinfo {author} {\bibfnamefont {Marc~G}\ \bibnamefont
  {Bellemare}}, \bibinfo {author} {\bibfnamefont {Yavar}\ \bibnamefont
  {Naddaf}}, \bibinfo {author} {\bibfnamefont {Joel}\ \bibnamefont {Veness}}, \
  and\ \bibinfo {author} {\bibfnamefont {Michael}\ \bibnamefont {Bowling}},\
  }\bibfield  {title} {\enquote {\bibinfo {title} {The arcade learning
  environment: An evaluation platform for general agents},}\ }\href@noop {}
  {\bibfield  {journal} {\bibinfo  {journal} {Journal of Artificial
  Intelligence Research}\ }\textbf {\bibinfo {volume} {47}},\ \bibinfo {pages}
  {253--279} (\bibinfo {year} {2013})}\BibitemShut {NoStop}%
\bibitem [{\citenamefont {Mnih}\ \emph {et~al.}(2016)\citenamefont {Mnih},
  \citenamefont {Badia}, \citenamefont {Mirza}, \citenamefont {Graves},
  \citenamefont {Lillicrap}, \citenamefont {Harley}, \citenamefont {Silver},\
  and\ \citenamefont {Kavukcuoglu}}]{Actor}%
  \BibitemOpen
  \bibfield  {author} {\bibinfo {author} {\bibfnamefont {Volodymyr}\
  \bibnamefont {Mnih}}, \bibinfo {author} {\bibfnamefont {Adria~Puigdomenech}\
  \bibnamefont {Badia}}, \bibinfo {author} {\bibfnamefont {Mehdi}\ \bibnamefont
  {Mirza}}, \bibinfo {author} {\bibfnamefont {Alex}\ \bibnamefont {Graves}},
  \bibinfo {author} {\bibfnamefont {Timothy}\ \bibnamefont {Lillicrap}},
  \bibinfo {author} {\bibfnamefont {Tim}\ \bibnamefont {Harley}}, \bibinfo
  {author} {\bibfnamefont {David}\ \bibnamefont {Silver}}, \ and\ \bibinfo
  {author} {\bibfnamefont {Koray}\ \bibnamefont {Kavukcuoglu}},\ }\bibfield
  {title} {\enquote {\bibinfo {title} {Asynchronous methods for deep
  reinforcement learning},}\ }in\ \href@noop {} {\emph {\bibinfo {booktitle}
  {International conference on machine learning}}}\ (\bibinfo {year} {2016})\
  pp.\ \bibinfo {pages} {1928--1937}\BibitemShut {NoStop}%
\bibitem [{\citenamefont {Tassa}\ \emph {et~al.}(2018)\citenamefont {Tassa},
  \citenamefont {Doron}, \citenamefont {Muldal}, \citenamefont {Erez},
  \citenamefont {Li}, \citenamefont {Casas}, \citenamefont {Budden},
  \citenamefont {Abdolmaleki}, \citenamefont {Merel}, \citenamefont {Lefrancq}
  \emph {et~al.}}]{DeepMind}%
  \BibitemOpen
  \bibfield  {author} {\bibinfo {author} {\bibfnamefont {Yuval}\ \bibnamefont
  {Tassa}}, \bibinfo {author} {\bibfnamefont {Yotam}\ \bibnamefont {Doron}},
  \bibinfo {author} {\bibfnamefont {Alistair}\ \bibnamefont {Muldal}}, \bibinfo
  {author} {\bibfnamefont {Tom}\ \bibnamefont {Erez}}, \bibinfo {author}
  {\bibfnamefont {Yazhe}\ \bibnamefont {Li}}, \bibinfo {author} {\bibfnamefont
  {Diego de~Las}\ \bibnamefont {Casas}}, \bibinfo {author} {\bibfnamefont
  {David}\ \bibnamefont {Budden}}, \bibinfo {author} {\bibfnamefont {Abbas}\
  \bibnamefont {Abdolmaleki}}, \bibinfo {author} {\bibfnamefont {Josh}\
  \bibnamefont {Merel}}, \bibinfo {author} {\bibfnamefont {Andrew}\
  \bibnamefont {Lefrancq}},  \emph {et~al.},\ }\bibfield  {title} {\enquote
  {\bibinfo {title} {Deepmind control suite},}\ }\href@noop {} {\bibfield
  {journal} {\bibinfo  {journal} {arXiv preprint arXiv:1801.00690}\ } (\bibinfo
  {year} {2018})}\BibitemShut {NoStop}%
\bibitem [{\citenamefont {Todorov}\ \emph {et~al.}(2012)\citenamefont
  {Todorov}, \citenamefont {Erez},\ and\ \citenamefont {Tassa}}]{MuJoCo}%
  \BibitemOpen
  \bibfield  {author} {\bibinfo {author} {\bibfnamefont {Emanuel}\ \bibnamefont
  {Todorov}}, \bibinfo {author} {\bibfnamefont {Tom}\ \bibnamefont {Erez}}, \
  and\ \bibinfo {author} {\bibfnamefont {Yuval}\ \bibnamefont {Tassa}},\
  }\bibfield  {title} {\enquote {\bibinfo {title} {Mujoco: A physics engine for
  model-based control},}\ }in\ \href@noop {} {\emph {\bibinfo {booktitle}
  {Intelligent Robots and Systems (IROS), 2012 IEEE/RSJ International
  Conference on}}}\ (\bibinfo {organization} {IEEE},\ \bibinfo {year} {2012})\
  pp.\ \bibinfo {pages} {5026--5033}\BibitemShut {NoStop}%
\bibitem [{\citenamefont {Puterman}(2014)}]{MDP}%
  \BibitemOpen
  \bibfield  {author} {\bibinfo {author} {\bibfnamefont {Martin~L}\
  \bibnamefont {Puterman}},\ }\href@noop {} {\emph {\bibinfo {title} {Markov
  decision processes: discrete stochastic dynamic programming}}}\ (\bibinfo
  {publisher} {John Wiley \& Sons},\ \bibinfo {year} {2014})\BibitemShut
  {NoStop}%
\bibitem [{\citenamefont {Asperti}\ \emph {et~al.}(2018)\citenamefont
  {Asperti}, \citenamefont {Cortesi},\ and\ \citenamefont {Sovrano}}]{Rogue}%
  \BibitemOpen
  \bibfield  {author} {\bibinfo {author} {\bibfnamefont {Andrea}\ \bibnamefont
  {Asperti}}, \bibinfo {author} {\bibfnamefont {Daniele}\ \bibnamefont
  {Cortesi}}, \ and\ \bibinfo {author} {\bibfnamefont {Francesco}\ \bibnamefont
  {Sovrano}},\ }\bibfield  {title} {\enquote {\bibinfo {title} {Crawling in
  rogue's dungeons with (partitioned) a3c},}\ }\href@noop {} {\bibfield
  {journal} {\bibinfo  {journal} {arXiv preprint arXiv:1804.08685}\ } (\bibinfo
  {year} {2018})}\BibitemShut {NoStop}%
\bibitem [{\citenamefont {Riedmiller}(2005)}]{NFQ}%
  \BibitemOpen
  \bibfield  {author} {\bibinfo {author} {\bibfnamefont {Martin}\ \bibnamefont
  {Riedmiller}},\ }\bibfield  {title} {\enquote {\bibinfo {title} {Neural
  fitted q iteration--first experiences with a data efficient neural
  reinforcement learning method},}\ }in\ \href@noop {} {\emph {\bibinfo
  {booktitle} {European Conference on Machine Learning}}}\ (\bibinfo
  {organization} {Springer},\ \bibinfo {year} {2005})\ pp.\ \bibinfo {pages}
  {317--328}\BibitemShut {NoStop}%
\bibitem [{\citenamefont {Riedmiller}\ \emph {et~al.}(2009)\citenamefont
  {Riedmiller}, \citenamefont {Gabel}, \citenamefont {Hafner},\ and\
  \citenamefont {Lange}}]{Soccer}%
  \BibitemOpen
  \bibfield  {author} {\bibinfo {author} {\bibfnamefont {Martin}\ \bibnamefont
  {Riedmiller}}, \bibinfo {author} {\bibfnamefont {Thomas}\ \bibnamefont
  {Gabel}}, \bibinfo {author} {\bibfnamefont {Roland}\ \bibnamefont {Hafner}},
  \ and\ \bibinfo {author} {\bibfnamefont {Sascha}\ \bibnamefont {Lange}},\
  }\bibfield  {title} {\enquote {\bibinfo {title} {Reinforcement learning for
  robot soccer},}\ }\href@noop {} {\bibfield  {journal} {\bibinfo  {journal}
  {Autonomous Robots}\ }\textbf {\bibinfo {volume} {27}},\ \bibinfo {pages}
  {55--73} (\bibinfo {year} {2009})}\BibitemShut {NoStop}%
\bibitem [{\citenamefont {Schulman}\ \emph {et~al.}(2017)\citenamefont
  {Schulman}, \citenamefont {Wolski}, \citenamefont {Dhariwal}, \citenamefont
  {Radford},\ and\ \citenamefont {Klimov}}]{PPO}%
  \BibitemOpen
  \bibfield  {author} {\bibinfo {author} {\bibfnamefont {John}\ \bibnamefont
  {Schulman}}, \bibinfo {author} {\bibfnamefont {Filip}\ \bibnamefont
  {Wolski}}, \bibinfo {author} {\bibfnamefont {Prafulla}\ \bibnamefont
  {Dhariwal}}, \bibinfo {author} {\bibfnamefont {Alec}\ \bibnamefont
  {Radford}}, \ and\ \bibinfo {author} {\bibfnamefont {Oleg}\ \bibnamefont
  {Klimov}},\ }\bibfield  {title} {\enquote {\bibinfo {title} {Proximal policy
  optimization algorithms},}\ }\href@noop {} {\bibfield  {journal} {\bibinfo
  {journal} {arXiv preprint arXiv:1707.06347}\ } (\bibinfo {year}
  {2017})}\BibitemShut {NoStop}%
\bibitem [{\citenamefont {Brockman}\ \emph {et~al.}(2016)\citenamefont
  {Brockman}, \citenamefont {Cheung}, \citenamefont {Pettersson}, \citenamefont
  {Schneider}, \citenamefont {Schulman}, \citenamefont {Tang},\ and\
  \citenamefont {Zaremba}}]{OpenAI}%
  \BibitemOpen
  \bibfield  {author} {\bibinfo {author} {\bibfnamefont {Greg}\ \bibnamefont
  {Brockman}}, \bibinfo {author} {\bibfnamefont {Vicki}\ \bibnamefont
  {Cheung}}, \bibinfo {author} {\bibfnamefont {Ludwig}\ \bibnamefont
  {Pettersson}}, \bibinfo {author} {\bibfnamefont {Jonas}\ \bibnamefont
  {Schneider}}, \bibinfo {author} {\bibfnamefont {John}\ \bibnamefont
  {Schulman}}, \bibinfo {author} {\bibfnamefont {Jie}\ \bibnamefont {Tang}}, \
  and\ \bibinfo {author} {\bibfnamefont {Wojciech}\ \bibnamefont {Zaremba}},\
  }\bibfield  {title} {\enquote {\bibinfo {title} {Openai gym},}\ }\href@noop
  {} {\bibfield  {journal} {\bibinfo  {journal} {arXiv preprint
  arXiv:1606.01540}\ } (\bibinfo {year} {2016})}\BibitemShut {NoStop}%
\bibitem [{\citenamefont {Kempka}\ \emph {et~al.}(2016)\citenamefont {Kempka},
  \citenamefont {Wydmuch}, \citenamefont {Runc}, \citenamefont {Toczek},\ and\
  \citenamefont {Ja{\'s}kowski}}]{VizDoom}%
  \BibitemOpen
  \bibfield  {author} {\bibinfo {author} {\bibfnamefont {Micha{\l}}\
  \bibnamefont {Kempka}}, \bibinfo {author} {\bibfnamefont {Marek}\
  \bibnamefont {Wydmuch}}, \bibinfo {author} {\bibfnamefont {Grzegorz}\
  \bibnamefont {Runc}}, \bibinfo {author} {\bibfnamefont {Jakub}\ \bibnamefont
  {Toczek}}, \ and\ \bibinfo {author} {\bibfnamefont {Wojciech}\ \bibnamefont
  {Ja{\'s}kowski}},\ }\bibfield  {title} {\enquote {\bibinfo {title} {Vizdoom:
  A doom-based ai research platform for visual reinforcement learning},}\ }in\
  \href@noop {} {\emph {\bibinfo {booktitle} {Computational Intelligence and
  Games (CIG), 2016 IEEE Conference on}}}\ (\bibinfo {organization} {IEEE},\
  \bibinfo {year} {2016})\ pp.\ \bibinfo {pages} {1--8}\BibitemShut {NoStop}%
\bibitem [{\citenamefont {Wydmuch}\ \emph {et~al.}(2018)\citenamefont
  {Wydmuch}, \citenamefont {Kempka},\ and\ \citenamefont
  {Ja{\'s}kowski}}]{VizDoom2}%
  \BibitemOpen
  \bibfield  {author} {\bibinfo {author} {\bibfnamefont {Marek}\ \bibnamefont
  {Wydmuch}}, \bibinfo {author} {\bibfnamefont {Micha{\l}}\ \bibnamefont
  {Kempka}}, \ and\ \bibinfo {author} {\bibfnamefont {Wojciech}\ \bibnamefont
  {Ja{\'s}kowski}},\ }\bibfield  {title} {\enquote {\bibinfo {title} {Vizdoom
  competitions: Playing doom from pixels},}\ }\href@noop {} {\bibfield
  {journal} {\bibinfo  {journal} {arXiv preprint arXiv:1809.03470}\ } (\bibinfo
  {year} {2018})}\BibitemShut {NoStop}%
\bibitem [{\citenamefont {Silver}\ \emph {et~al.}(2016)\citenamefont {Silver},
  \citenamefont {Huang}, \citenamefont {Maddison}, \citenamefont {Guez},
  \citenamefont {Sifre}, \citenamefont {Van Den~Driessche}, \citenamefont
  {Schrittwieser}, \citenamefont {Antonoglou}, \citenamefont {Panneershelvam},
  \citenamefont {Lanctot} \emph {et~al.}}]{Go}%
  \BibitemOpen
  \bibfield  {author} {\bibinfo {author} {\bibfnamefont {David}\ \bibnamefont
  {Silver}}, \bibinfo {author} {\bibfnamefont {Aja}\ \bibnamefont {Huang}},
  \bibinfo {author} {\bibfnamefont {Chris~J}\ \bibnamefont {Maddison}},
  \bibinfo {author} {\bibfnamefont {Arthur}\ \bibnamefont {Guez}}, \bibinfo
  {author} {\bibfnamefont {Laurent}\ \bibnamefont {Sifre}}, \bibinfo {author}
  {\bibfnamefont {George}\ \bibnamefont {Van Den~Driessche}}, \bibinfo {author}
  {\bibfnamefont {Julian}\ \bibnamefont {Schrittwieser}}, \bibinfo {author}
  {\bibfnamefont {Ioannis}\ \bibnamefont {Antonoglou}}, \bibinfo {author}
  {\bibfnamefont {Veda}\ \bibnamefont {Panneershelvam}}, \bibinfo {author}
  {\bibfnamefont {Marc}\ \bibnamefont {Lanctot}},  \emph {et~al.},\ }\bibfield
  {title} {\enquote {\bibinfo {title} {Mastering the game of go with deep
  neural networks and tree search},}\ }\href@noop {} {\bibfield  {journal}
  {\bibinfo  {journal} {nature}\ }\textbf {\bibinfo {volume} {529}},\ \bibinfo
  {pages} {484} (\bibinfo {year} {2016})}\BibitemShut {NoStop}%
\bibitem [{\citenamefont {Silver}\ \emph {et~al.}(2017)\citenamefont {Silver},
  \citenamefont {Schrittwieser}, \citenamefont {Simonyan}, \citenamefont
  {Antonoglou}, \citenamefont {Huang}, \citenamefont {Guez}, \citenamefont
  {Hubert}, \citenamefont {Baker}, \citenamefont {Lai}, \citenamefont {Bolton}
  \emph {et~al.}}]{Go2}%
  \BibitemOpen
  \bibfield  {author} {\bibinfo {author} {\bibfnamefont {David}\ \bibnamefont
  {Silver}}, \bibinfo {author} {\bibfnamefont {Julian}\ \bibnamefont
  {Schrittwieser}}, \bibinfo {author} {\bibfnamefont {Karen}\ \bibnamefont
  {Simonyan}}, \bibinfo {author} {\bibfnamefont {Ioannis}\ \bibnamefont
  {Antonoglou}}, \bibinfo {author} {\bibfnamefont {Aja}\ \bibnamefont {Huang}},
  \bibinfo {author} {\bibfnamefont {Arthur}\ \bibnamefont {Guez}}, \bibinfo
  {author} {\bibfnamefont {Thomas}\ \bibnamefont {Hubert}}, \bibinfo {author}
  {\bibfnamefont {Lucas}\ \bibnamefont {Baker}}, \bibinfo {author}
  {\bibfnamefont {Matthew}\ \bibnamefont {Lai}}, \bibinfo {author}
  {\bibfnamefont {Adrian}\ \bibnamefont {Bolton}},  \emph {et~al.},\ }\bibfield
   {title} {\enquote {\bibinfo {title} {Mastering the game of go without human
  knowledge},}\ }\href@noop {} {\bibfield  {journal} {\bibinfo  {journal}
  {Nature}\ }\textbf {\bibinfo {volume} {550}},\ \bibinfo {pages} {354}
  (\bibinfo {year} {2017})}\BibitemShut {NoStop}%
\bibitem [{\citenamefont {Such}\ \emph {et~al.}(2017)\citenamefont {Such},
  \citenamefont {Madhavan}, \citenamefont {Conti}, \citenamefont {Lehman},
  \citenamefont {Stanley},\ and\ \citenamefont {Clune}}]{Guber}%
  \BibitemOpen
  \bibfield  {author} {\bibinfo {author} {\bibfnamefont {Felipe~Petroski}\
  \bibnamefont {Such}}, \bibinfo {author} {\bibfnamefont {Vashisht}\
  \bibnamefont {Madhavan}}, \bibinfo {author} {\bibfnamefont {Edoardo}\
  \bibnamefont {Conti}}, \bibinfo {author} {\bibfnamefont {Joel}\ \bibnamefont
  {Lehman}}, \bibinfo {author} {\bibfnamefont {Kenneth~O}\ \bibnamefont
  {Stanley}}, \ and\ \bibinfo {author} {\bibfnamefont {Jeff}\ \bibnamefont
  {Clune}},\ }\bibfield  {title} {\enquote {\bibinfo {title} {Deep
  neuroevolution: genetic algorithms are a competitive alternative for training
  deep neural networks for reinforcement learning},}\ }\href@noop {} {\bibfield
   {journal} {\bibinfo  {journal} {arXiv preprint arXiv:1712.06567}\ }
  (\bibinfo {year} {2017})}\BibitemShut {NoStop}%
\bibitem [{\citenamefont {Holland}(1992)}]{GA}%
  \BibitemOpen
  \bibfield  {author} {\bibinfo {author} {\bibfnamefont {John~H}\ \bibnamefont
  {Holland}},\ }\bibfield  {title} {\enquote {\bibinfo {title} {Genetic
  algorithms},}\ }\href@noop {} {\bibfield  {journal} {\bibinfo  {journal}
  {Scientific american}\ }\textbf {\bibinfo {volume} {267}},\ \bibinfo {pages}
  {66--73} (\bibinfo {year} {1992})}\BibitemShut {NoStop}%
\bibitem [{\citenamefont {Fogel}\ and\ \citenamefont {Stayton}(1994)}]{GA2}%
  \BibitemOpen
  \bibfield  {author} {\bibinfo {author} {\bibfnamefont {David~B}\ \bibnamefont
  {Fogel}}\ and\ \bibinfo {author} {\bibfnamefont {Lauren~C}\ \bibnamefont
  {Stayton}},\ }\bibfield  {title} {\enquote {\bibinfo {title} {On the
  effectiveness of crossover in simulated evolutionary optimization},}\
  }\href@noop {} {\bibfield  {journal} {\bibinfo  {journal} {BioSystems}\
  }\textbf {\bibinfo {volume} {32}},\ \bibinfo {pages} {171--182} (\bibinfo
  {year} {1994})}\BibitemShut {NoStop}%
\bibitem [{\citenamefont {Lehman}\ \emph
  {et~al.}(2018{\natexlab{a}})\citenamefont {Lehman}, \citenamefont {Chen},
  \citenamefont {Clune},\ and\ \citenamefont {Stanley}}]{lehman2018more}%
  \BibitemOpen
  \bibfield  {author} {\bibinfo {author} {\bibfnamefont {Joel}\ \bibnamefont
  {Lehman}}, \bibinfo {author} {\bibfnamefont {Jay}\ \bibnamefont {Chen}},
  \bibinfo {author} {\bibfnamefont {Jeff}\ \bibnamefont {Clune}}, \ and\
  \bibinfo {author} {\bibfnamefont {Kenneth~O}\ \bibnamefont {Stanley}},\
  }\bibfield  {title} {\enquote {\bibinfo {title} {Es is more than just a
  traditional finite-difference approximator},}\ }in\ \href@noop {} {\emph
  {\bibinfo {booktitle} {Proceedings of the Genetic and Evolutionary
  Computation Conference}}}\ (\bibinfo {year} {2018})\ pp.\ \bibinfo {pages}
  {450--457}\BibitemShut {NoStop}%
\bibitem [{\citenamefont {Salimans}\ \emph {et~al.}(2017)\citenamefont
  {Salimans}, \citenamefont {Ho}, \citenamefont {Chen}, \citenamefont {Sidor},\
  and\ \citenamefont {Sutskever}}]{salimans2017evolution}%
  \BibitemOpen
  \bibfield  {author} {\bibinfo {author} {\bibfnamefont {Tim}\ \bibnamefont
  {Salimans}}, \bibinfo {author} {\bibfnamefont {Jonathan}\ \bibnamefont {Ho}},
  \bibinfo {author} {\bibfnamefont {Xi}~\bibnamefont {Chen}}, \bibinfo {author}
  {\bibfnamefont {Szymon}\ \bibnamefont {Sidor}}, \ and\ \bibinfo {author}
  {\bibfnamefont {Ilya}\ \bibnamefont {Sutskever}},\ }\bibfield  {title}
  {\enquote {\bibinfo {title} {Evolution strategies as a scalable alternative
  to reinforcement learning},}\ }\href@noop {} {\bibfield  {journal} {\bibinfo
  {journal} {arXiv preprint arXiv:1703.03864}\ } (\bibinfo {year}
  {2017})}\BibitemShut {NoStop}%
\bibitem [{\citenamefont {Zhang}\ \emph {et~al.}(2017)\citenamefont {Zhang},
  \citenamefont {Clune},\ and\ \citenamefont
  {Stanley}}]{zhang2017relationship}%
  \BibitemOpen
  \bibfield  {author} {\bibinfo {author} {\bibfnamefont {Xingwen}\ \bibnamefont
  {Zhang}}, \bibinfo {author} {\bibfnamefont {Jeff}\ \bibnamefont {Clune}}, \
  and\ \bibinfo {author} {\bibfnamefont {Kenneth~O}\ \bibnamefont {Stanley}},\
  }\bibfield  {title} {\enquote {\bibinfo {title} {On the relationship between
  the openai evolution strategy and stochastic gradient descent},}\ }\href@noop
  {} {\bibfield  {journal} {\bibinfo  {journal} {arXiv preprint
  arXiv:1712.06564}\ } (\bibinfo {year} {2017})}\BibitemShut {NoStop}%
\bibitem [{\citenamefont {Lehman}\ \emph
  {et~al.}(2018{\natexlab{b}})\citenamefont {Lehman}, \citenamefont {Chen},
  \citenamefont {Clune},\ and\ \citenamefont {Stanley}}]{lehman2018safe}%
  \BibitemOpen
  \bibfield  {author} {\bibinfo {author} {\bibfnamefont {Joel}\ \bibnamefont
  {Lehman}}, \bibinfo {author} {\bibfnamefont {Jay}\ \bibnamefont {Chen}},
  \bibinfo {author} {\bibfnamefont {Jeff}\ \bibnamefont {Clune}}, \ and\
  \bibinfo {author} {\bibfnamefont {Kenneth~O}\ \bibnamefont {Stanley}},\
  }\bibfield  {title} {\enquote {\bibinfo {title} {Safe mutations for deep and
  recurrent neural networks through output gradients},}\ }in\ \href@noop {}
  {\emph {\bibinfo {booktitle} {Proceedings of the Genetic and Evolutionary
  Computation Conference}}}\ (\bibinfo {year} {2018})\ pp.\ \bibinfo {pages}
  {117--124}\BibitemShut {NoStop}%
\bibitem [{\citenamefont {Conti}\ \emph {et~al.}(2018)\citenamefont {Conti},
  \citenamefont {Madhavan}, \citenamefont {Such}, \citenamefont {Lehman},
  \citenamefont {Stanley},\ and\ \citenamefont {Clune}}]{conti2018improving}%
  \BibitemOpen
  \bibfield  {author} {\bibinfo {author} {\bibfnamefont {Edoardo}\ \bibnamefont
  {Conti}}, \bibinfo {author} {\bibfnamefont {Vashisht}\ \bibnamefont
  {Madhavan}}, \bibinfo {author} {\bibfnamefont {Felipe~Petroski}\ \bibnamefont
  {Such}}, \bibinfo {author} {\bibfnamefont {Joel}\ \bibnamefont {Lehman}},
  \bibinfo {author} {\bibfnamefont {Kenneth}\ \bibnamefont {Stanley}}, \ and\
  \bibinfo {author} {\bibfnamefont {Jeff}\ \bibnamefont {Clune}},\ }\bibfield
  {title} {\enquote {\bibinfo {title} {Improving exploration in evolution
  strategies for deep reinforcement learning via a population of
  novelty-seeking agents},}\ }in\ \href@noop {} {\emph {\bibinfo {booktitle}
  {Advances in neural information processing systems}}}\ (\bibinfo {year}
  {2018})\ pp.\ \bibinfo {pages} {5027--5038}\BibitemShut {NoStop}%
\bibitem [{\citenamefont {Zhang}\ and\ \citenamefont
  {Glotzer}(2004)}]{zhang2004self}%
  \BibitemOpen
  \bibfield  {author} {\bibinfo {author} {\bibfnamefont {Zhenli}\ \bibnamefont
  {Zhang}}\ and\ \bibinfo {author} {\bibfnamefont {Sharon~C}\ \bibnamefont
  {Glotzer}},\ }\bibfield  {title} {\enquote {\bibinfo {title} {Self-assembly
  of patchy particles},}\ }\href@noop {} {\bibfield  {journal} {\bibinfo
  {journal} {Nano Letters}\ }\textbf {\bibinfo {volume} {4}},\ \bibinfo {pages}
  {1407--1413} (\bibinfo {year} {2004})}\BibitemShut {NoStop}%
\bibitem [{\citenamefont {Romano}\ \emph {et~al.}(2011)\citenamefont {Romano},
  \citenamefont {Sanz},\ and\ \citenamefont
  {Sciortino}}]{romano2011crystallization}%
  \BibitemOpen
  \bibfield  {author} {\bibinfo {author} {\bibfnamefont {Flavio}\ \bibnamefont
  {Romano}}, \bibinfo {author} {\bibfnamefont {Eduardo}\ \bibnamefont {Sanz}},
  \ and\ \bibinfo {author} {\bibfnamefont {Francesco}\ \bibnamefont
  {Sciortino}},\ }\bibfield  {title} {\enquote {\bibinfo {title}
  {Crystallization of tetrahedral patchy particles in silico},}\ }\href@noop {}
  {\bibfield  {journal} {\bibinfo  {journal} {The Journal of Chemical Physics}\
  }\textbf {\bibinfo {volume} {134}},\ \bibinfo {pages} {174502} (\bibinfo
  {year} {2011})}\BibitemShut {NoStop}%
\bibitem [{\citenamefont {Sciortino}\ \emph {et~al.}(2007)\citenamefont
  {Sciortino}, \citenamefont {Bianchi}, \citenamefont {Douglas},\ and\
  \citenamefont {Tartaglia}}]{sciortino2007self}%
  \BibitemOpen
  \bibfield  {author} {\bibinfo {author} {\bibfnamefont {Francesco}\
  \bibnamefont {Sciortino}}, \bibinfo {author} {\bibfnamefont {Emanuela}\
  \bibnamefont {Bianchi}}, \bibinfo {author} {\bibfnamefont {Jack~F}\
  \bibnamefont {Douglas}}, \ and\ \bibinfo {author} {\bibfnamefont {Piero}\
  \bibnamefont {Tartaglia}},\ }\bibfield  {title} {\enquote {\bibinfo {title}
  {Self-assembly of patchy particles into polymer chains: A parameter-free
  comparison between wertheim theory and monte carlo simulation},}\ }\href@noop
  {} {\bibfield  {journal} {\bibinfo  {journal} {The Journal of Chemical
  Physics}\ }\textbf {\bibinfo {volume} {126}},\ \bibinfo {pages} {194903}
  (\bibinfo {year} {2007})}\BibitemShut {NoStop}%
\bibitem [{\citenamefont {Doye}\ \emph
  {et~al.}(2007{\natexlab{b}})\citenamefont {Doye}, \citenamefont {Louis},
  \citenamefont {Lin}, \citenamefont {Allen}, \citenamefont {Noya},
  \citenamefont {Wilber}, \citenamefont {Kok},\ and\ \citenamefont
  {Lyus}}]{doye2007controlling}%
  \BibitemOpen
  \bibfield  {author} {\bibinfo {author} {\bibfnamefont {J.P.K.}\ \bibnamefont
  {Doye}}, \bibinfo {author} {\bibfnamefont {A.A.}\ \bibnamefont {Louis}},
  \bibinfo {author} {\bibfnamefont {I.C.}\ \bibnamefont {Lin}}, \bibinfo
  {author} {\bibfnamefont {L.R.}\ \bibnamefont {Allen}}, \bibinfo {author}
  {\bibfnamefont {E.G.}\ \bibnamefont {Noya}}, \bibinfo {author} {\bibfnamefont
  {A.W.}\ \bibnamefont {Wilber}}, \bibinfo {author} {\bibfnamefont {H.C.}\
  \bibnamefont {Kok}}, \ and\ \bibinfo {author} {\bibfnamefont
  {R.}~\bibnamefont {Lyus}},\ }\bibfield  {title} {\enquote {\bibinfo {title}
  {{Controlling crystallization and its absence: proteins, colloids and patchy
  models}},}\ }\href@noop {} {\bibfield  {journal} {\bibinfo  {journal}
  {Physical Chemistry Chemical Physics}\ }\textbf {\bibinfo {volume} {9}},\
  \bibinfo {pages} {2197--2205} (\bibinfo {year}
  {2007}{\natexlab{b}})}\BibitemShut {NoStop}%
\bibitem [{\citenamefont {Bianchi}\ \emph {et~al.}(2008)\citenamefont
  {Bianchi}, \citenamefont {Tartaglia}, \citenamefont {Zaccarelli},\ and\
  \citenamefont {Sciortino}}]{bianchi2008theoretical}%
  \BibitemOpen
  \bibfield  {author} {\bibinfo {author} {\bibfnamefont {Emanuela}\
  \bibnamefont {Bianchi}}, \bibinfo {author} {\bibfnamefont {Piero}\
  \bibnamefont {Tartaglia}}, \bibinfo {author} {\bibfnamefont {Emanuela}\
  \bibnamefont {Zaccarelli}}, \ and\ \bibinfo {author} {\bibfnamefont
  {Francesco}\ \bibnamefont {Sciortino}},\ }\bibfield  {title} {\enquote
  {\bibinfo {title} {Theoretical and numerical study of the phase diagram of
  patchy colloids: Ordered and disordered patch arrangements},}\ }\href@noop {}
  {\bibfield  {journal} {\bibinfo  {journal} {The Journal of Chemical Physics}\
  }\textbf {\bibinfo {volume} {128}},\ \bibinfo {pages} {144504} (\bibinfo
  {year} {2008})}\BibitemShut {NoStop}%
\bibitem [{\citenamefont {Doppelbauer}\ \emph {et~al.}(2010)\citenamefont
  {Doppelbauer}, \citenamefont {Bianchi},\ and\ \citenamefont
  {Kahl}}]{doppelbauer2010self}%
  \BibitemOpen
  \bibfield  {author} {\bibinfo {author} {\bibfnamefont {G{\"u}nther}\
  \bibnamefont {Doppelbauer}}, \bibinfo {author} {\bibfnamefont {Emanuela}\
  \bibnamefont {Bianchi}}, \ and\ \bibinfo {author} {\bibfnamefont {Gerhard}\
  \bibnamefont {Kahl}},\ }\bibfield  {title} {\enquote {\bibinfo {title}
  {Self-assembly scenarios of patchy colloidal particles in two dimensions},}\
  }\href@noop {} {\bibfield  {journal} {\bibinfo  {journal} {Journal of
  Physics: Condensed Matter}\ }\textbf {\bibinfo {volume} {22}},\ \bibinfo
  {pages} {104105} (\bibinfo {year} {2010})}\BibitemShut {NoStop}%
\bibitem [{\citenamefont {Whitelam}\ \emph
  {et~al.}(2014{\natexlab{b}})\citenamefont {Whitelam}, \citenamefont
  {Tamblyn}, \citenamefont {Haxton}, \citenamefont {Wieland}, \citenamefont
  {Champness}, \citenamefont {Garrahan},\ and\ \citenamefont
  {Beton}}]{whitelam2014common}%
  \BibitemOpen
  \bibfield  {author} {\bibinfo {author} {\bibfnamefont {Stephen}\ \bibnamefont
  {Whitelam}}, \bibinfo {author} {\bibfnamefont {Isaac}\ \bibnamefont
  {Tamblyn}}, \bibinfo {author} {\bibfnamefont {Thomas~K}\ \bibnamefont
  {Haxton}}, \bibinfo {author} {\bibfnamefont {Maria~B}\ \bibnamefont
  {Wieland}}, \bibinfo {author} {\bibfnamefont {Neil~R}\ \bibnamefont
  {Champness}}, \bibinfo {author} {\bibfnamefont {Juan~P}\ \bibnamefont
  {Garrahan}}, \ and\ \bibinfo {author} {\bibfnamefont {Peter~H}\ \bibnamefont
  {Beton}},\ }\bibfield  {title} {\enquote {\bibinfo {title} {Common physical
  framework explains phase behavior and dynamics of atomic, molecular, and
  polymeric network formers},}\ }\href@noop {} {\bibfield  {journal} {\bibinfo
  {journal} {Physical Review X}\ }\textbf {\bibinfo {volume} {4}},\ \bibinfo
  {pages} {011044} (\bibinfo {year} {2014}{\natexlab{b}})}\BibitemShut
  {NoStop}%
\bibitem [{\citenamefont {Duguet}\ \emph {et~al.}(2016)\citenamefont {Duguet},
  \citenamefont {Hubert}, \citenamefont {Chomette}, \citenamefont {Perro},\
  and\ \citenamefont {Ravaine}}]{duguet2016patchy}%
  \BibitemOpen
  \bibfield  {author} {\bibinfo {author} {\bibfnamefont {{\'E}tienne}\
  \bibnamefont {Duguet}}, \bibinfo {author} {\bibfnamefont {C{\'e}line}\
  \bibnamefont {Hubert}}, \bibinfo {author} {\bibfnamefont {Cyril}\
  \bibnamefont {Chomette}}, \bibinfo {author} {\bibfnamefont {Adeline}\
  \bibnamefont {Perro}}, \ and\ \bibinfo {author} {\bibfnamefont {Serge}\
  \bibnamefont {Ravaine}},\ }\bibfield  {title} {\enquote {\bibinfo {title}
  {Patchy colloidal particles for programmed self-assembly},}\ }\href@noop {}
  {\bibfield  {journal} {\bibinfo  {journal} {Comptes Rendus Chimie}\ }\textbf
  {\bibinfo {volume} {19}},\ \bibinfo {pages} {173--182} (\bibinfo {year}
  {2016})}\BibitemShut {NoStop}%
\bibitem [{\citenamefont {Frenkel}\ and\ \citenamefont
  {Smit}(1996)}]{frenkel1996understanding}%
  \BibitemOpen
  \bibfield  {author} {\bibinfo {author} {\bibfnamefont {D.}~\bibnamefont
  {Frenkel}}\ and\ \bibinfo {author} {\bibfnamefont {B.}~\bibnamefont {Smit}},\
  }\href@noop {} {\emph {\bibinfo {title} {{Understanding Molecular Simulation:
  From Algorithms to Applications}}}}\ (\bibinfo  {publisher} {Academic Press,
  Inc. Orlando, FL, USA},\ \bibinfo {year} {1996})\BibitemShut {NoStop}%
\bibitem [{\citenamefont {Klotsa}\ and\ \citenamefont
  {Jack}(2013)}]{klotsa2013controlling}%
  \BibitemOpen
  \bibfield  {author} {\bibinfo {author} {\bibfnamefont {Daphne}\ \bibnamefont
  {Klotsa}}\ and\ \bibinfo {author} {\bibfnamefont {Robert~L}\ \bibnamefont
  {Jack}},\ }\bibfield  {title} {\enquote {\bibinfo {title} {Controlling
  crystal self-assembly using a real-time feedback scheme},}\ }\href@noop {}
  {\bibfield  {journal} {\bibinfo  {journal} {The Journal of Chemical Physics}\
  }\textbf {\bibinfo {volume} {138}},\ \bibinfo {pages} {094502} (\bibinfo
  {year} {2013})}\BibitemShut {NoStop}%
\bibitem [{\citenamefont {Tang}\ \emph {et~al.}(2016)\citenamefont {Tang},
  \citenamefont {Rupp}, \citenamefont {Yang}, \citenamefont {Edwards},
  \citenamefont {Grover},\ and\ \citenamefont {Bevan}}]{tang2016optimal}%
  \BibitemOpen
  \bibfield  {author} {\bibinfo {author} {\bibfnamefont {Xun}\ \bibnamefont
  {Tang}}, \bibinfo {author} {\bibfnamefont {Bradley}\ \bibnamefont {Rupp}},
  \bibinfo {author} {\bibfnamefont {Yuguang}\ \bibnamefont {Yang}}, \bibinfo
  {author} {\bibfnamefont {Tara~D}\ \bibnamefont {Edwards}}, \bibinfo {author}
  {\bibfnamefont {Martha~A}\ \bibnamefont {Grover}}, \ and\ \bibinfo {author}
  {\bibfnamefont {Michael~A}\ \bibnamefont {Bevan}},\ }\bibfield  {title}
  {\enquote {\bibinfo {title} {Optimal feedback controlled assembly of perfect
  crystals},}\ }\href@noop {} {\bibfield  {journal} {\bibinfo  {journal} {ACS
  nano}\ }\textbf {\bibinfo {volume} {10}},\ \bibinfo {pages} {6791--6798}
  (\bibinfo {year} {2016})}\BibitemShut {NoStop}%
\bibitem [{\citenamefont {Miskin}\ \emph {et~al.}(2016)\citenamefont {Miskin},
  \citenamefont {Khaira}, \citenamefont {de~Pablo},\ and\ \citenamefont
  {Jaeger}}]{miskin2016turning}%
  \BibitemOpen
  \bibfield  {author} {\bibinfo {author} {\bibfnamefont {Marc~Z}\ \bibnamefont
  {Miskin}}, \bibinfo {author} {\bibfnamefont {Gurdaman}\ \bibnamefont
  {Khaira}}, \bibinfo {author} {\bibfnamefont {Juan~J}\ \bibnamefont
  {de~Pablo}}, \ and\ \bibinfo {author} {\bibfnamefont {Heinrich~M}\
  \bibnamefont {Jaeger}},\ }\bibfield  {title} {\enquote {\bibinfo {title}
  {Turning statistical physics models into materials design engines},}\
  }\href@noop {} {\bibfield  {journal} {\bibinfo  {journal} {Proceedings of the
  National Academy of Sciences}\ }\textbf {\bibinfo {volume} {113}},\ \bibinfo
  {pages} {34--39} (\bibinfo {year} {2016})}\BibitemShut {NoStop}%
\bibitem [{\citenamefont {Long}\ \emph {et~al.}(2015)\citenamefont {Long},
  \citenamefont {Zhang}, \citenamefont {Granick},\ and\ \citenamefont
  {Ferguson}}]{long2015machine}%
  \BibitemOpen
  \bibfield  {author} {\bibinfo {author} {\bibfnamefont {Andrew~W}\
  \bibnamefont {Long}}, \bibinfo {author} {\bibfnamefont {Jie}\ \bibnamefont
  {Zhang}}, \bibinfo {author} {\bibfnamefont {Steve}\ \bibnamefont {Granick}},
  \ and\ \bibinfo {author} {\bibfnamefont {Andrew~L}\ \bibnamefont
  {Ferguson}},\ }\bibfield  {title} {\enquote {\bibinfo {title} {Machine
  learning assembly landscapes from particle tracking data},}\ }\href@noop {}
  {\bibfield  {journal} {\bibinfo  {journal} {Soft Matter}\ }\textbf {\bibinfo
  {volume} {11}},\ \bibinfo {pages} {8141--8153} (\bibinfo {year}
  {2015})}\BibitemShut {NoStop}%
\bibitem [{\citenamefont {Long}\ and\ \citenamefont
  {Ferguson}(2014)}]{long2014nonlinear}%
  \BibitemOpen
  \bibfield  {author} {\bibinfo {author} {\bibfnamefont {Andrew~W}\
  \bibnamefont {Long}}\ and\ \bibinfo {author} {\bibfnamefont {Andrew~L}\
  \bibnamefont {Ferguson}},\ }\bibfield  {title} {\enquote {\bibinfo {title}
  {Nonlinear machine learning of patchy colloid self-assembly pathways and
  mechanisms},}\ }\href@noop {} {\bibfield  {journal} {\bibinfo  {journal} {The
  Journal of Physical Chemistry B}\ }\textbf {\bibinfo {volume} {118}},\
  \bibinfo {pages} {4228--4244} (\bibinfo {year} {2014})}\BibitemShut {NoStop}%
\bibitem [{\citenamefont {Lindquist}\ \emph {et~al.}(2016)\citenamefont
  {Lindquist}, \citenamefont {Jadrich},\ and\ \citenamefont
  {Truskett}}]{lindquist2016communication}%
  \BibitemOpen
  \bibfield  {author} {\bibinfo {author} {\bibfnamefont {Beth~A}\ \bibnamefont
  {Lindquist}}, \bibinfo {author} {\bibfnamefont {Ryan~B}\ \bibnamefont
  {Jadrich}}, \ and\ \bibinfo {author} {\bibfnamefont {Thomas~M}\ \bibnamefont
  {Truskett}},\ }\bibfield  {title} {\enquote {\bibinfo {title} {Communication:
  Inverse design for self-assembly via on-the-fly optimization},}\ }\href@noop
  {} {\bibfield  {journal} {\bibinfo  {journal} {Journal of Chemical Physics}\
  }\textbf {\bibinfo {volume} {145}} (\bibinfo {year} {2016})}\BibitemShut
  {NoStop}%
\bibitem [{\citenamefont {Thurston}\ and\ \citenamefont
  {Ferguson}(2018)}]{thurston2018machine}%
  \BibitemOpen
  \bibfield  {author} {\bibinfo {author} {\bibfnamefont {Bryce~A}\ \bibnamefont
  {Thurston}}\ and\ \bibinfo {author} {\bibfnamefont {Andrew~L}\ \bibnamefont
  {Ferguson}},\ }\bibfield  {title} {\enquote {\bibinfo {title} {Machine
  learning and molecular design of self-assembling-conjugated oligopeptides},}\
  }\href@noop {} {\bibfield  {journal} {\bibinfo  {journal} {Molecular
  Simulation}\ }\textbf {\bibinfo {volume} {44}},\ \bibinfo {pages} {930--945}
  (\bibinfo {year} {2018})}\BibitemShut {NoStop}%
\bibitem [{\citenamefont {Ferguson}(2017)}]{ferguson2017machine}%
  \BibitemOpen
  \bibfield  {author} {\bibinfo {author} {\bibfnamefont {Andrew~L}\
  \bibnamefont {Ferguson}},\ }\bibfield  {title} {\enquote {\bibinfo {title}
  {Machine learning and data science in soft materials engineering},}\
  }\href@noop {} {\bibfield  {journal} {\bibinfo  {journal} {Journal of
  Physics: Condensed Matter}\ }\textbf {\bibinfo {volume} {30}},\ \bibinfo
  {pages} {043002} (\bibinfo {year} {2017})}\BibitemShut {NoStop}%
\bibitem [{\citenamefont {Kern}\ and\ \citenamefont
  {Frenkel}(2003)}]{kern2003fluid}%
  \BibitemOpen
  \bibfield  {author} {\bibinfo {author} {\bibfnamefont {Norbert}\ \bibnamefont
  {Kern}}\ and\ \bibinfo {author} {\bibfnamefont {Daan}\ \bibnamefont
  {Frenkel}},\ }\bibfield  {title} {\enquote {\bibinfo {title} {Fluid--fluid
  coexistence in colloidal systems with short-ranged strongly directional
  attraction},}\ }\href@noop {} {\bibfield  {journal} {\bibinfo  {journal} {The
  Journal of Chemical Physics}\ }\textbf {\bibinfo {volume} {118}},\ \bibinfo
  {pages} {9882} (\bibinfo {year} {2003})}\BibitemShut {NoStop}%
\bibitem [{\citenamefont {Whitelam}(2016)}]{whitelam2016minimal}%
  \BibitemOpen
  \bibfield  {author} {\bibinfo {author} {\bibfnamefont {Stephen}\ \bibnamefont
  {Whitelam}},\ }\bibfield  {title} {\enquote {\bibinfo {title} {Minimal
  positive design for self-assembly of the archimedean tilings},}\ }\href@noop
  {} {\bibfield  {journal} {\bibinfo  {journal} {Physical Review Letters}\
  }\textbf {\bibinfo {volume} {117}},\ \bibinfo {pages} {228003} (\bibinfo
  {year} {2016})}\BibitemShut {NoStop}%
\bibitem [{\citenamefont {Whitelam}\ \emph {et~al.}(2009)\citenamefont
  {Whitelam}, \citenamefont {Feng}, \citenamefont {Hagan},\ and\ \citenamefont
  {Geissler}}]{whitelam2009role}%
  \BibitemOpen
  \bibfield  {author} {\bibinfo {author} {\bibfnamefont {Stephen}\ \bibnamefont
  {Whitelam}}, \bibinfo {author} {\bibfnamefont {Edward~H}\ \bibnamefont
  {Feng}}, \bibinfo {author} {\bibfnamefont {Michael~F}\ \bibnamefont {Hagan}},
  \ and\ \bibinfo {author} {\bibfnamefont {Phillip~L}\ \bibnamefont
  {Geissler}},\ }\bibfield  {title} {\enquote {\bibinfo {title} {The role of
  collective motion in examples of coarsening and self-assembly},}\ }\href@noop
  {} {\bibfield  {journal} {\bibinfo  {journal} {Soft Matter}\ }\textbf
  {\bibinfo {volume} {5}},\ \bibinfo {pages} {1251--1262} (\bibinfo {year}
  {2009})}\BibitemShut {NoStop}%
\bibitem [{\citenamefont {Hedges}()}]{VMMC_3}%
  \BibitemOpen
  \bibfield  {author} {\bibinfo {author} {\bibfnamefont {L.~O.}\ \bibnamefont
  {Hedges}},\ }\href {http://vmmc.xyz} {\enquote {\bibinfo {title}
  {http://vmmc.xyz},}\ }\BibitemShut {NoStop}%
\bibitem [{\citenamefont {Haxton}\ \emph {et~al.}(2015)\citenamefont {Haxton},
  \citenamefont {Hedges},\ and\ \citenamefont
  {Whitelam}}]{haxton2015crystallization}%
  \BibitemOpen
  \bibfield  {author} {\bibinfo {author} {\bibfnamefont {Thomas~K}\
  \bibnamefont {Haxton}}, \bibinfo {author} {\bibfnamefont {Lester~O}\
  \bibnamefont {Hedges}}, \ and\ \bibinfo {author} {\bibfnamefont {Stephen}\
  \bibnamefont {Whitelam}},\ }\bibfield  {title} {\enquote {\bibinfo {title}
  {Crystallization and arrest mechanisms of model colloids},}\ }\href@noop {}
  {\bibfield  {journal} {\bibinfo  {journal} {Soft matter}\ }\textbf {\bibinfo
  {volume} {11}},\ \bibinfo {pages} {9307--9320} (\bibinfo {year}
  {2015})}\BibitemShut {NoStop}%
\bibitem [{Note1()}]{Note1}%
  \BibitemOpen
  \bibinfo {note} {The natural way to measure ``real'' time in such a system is
  to advance the clock by an amount $(1+M)^{-1}$ upon making an attempted move.
  Dense systems and sparse systems then take very different amounts of CPU time
  to run. In order to move simulation generations efficiently through our
  computer cluster we instead updated the clock by one unit upon making a move.
  In this way we work in the constant event-number ensemble.}\BibitemShut
  {Stop}%
\bibitem [{\citenamefont {Whitelam}(2018)}]{whitelam2018strong}%
  \BibitemOpen
  \bibfield  {author} {\bibinfo {author} {\bibfnamefont {Stephen}\ \bibnamefont
  {Whitelam}},\ }\bibfield  {title} {\enquote {\bibinfo {title} {Strong bonds
  and far-from-equilibrium conditions minimize errors in lattice-gas growth},}\
  }\href@noop {} {\bibfield  {journal} {\bibinfo  {journal} {The Journal of
  Chemical Physics}\ }\textbf {\bibinfo {volume} {149}},\ \bibinfo {pages}
  {104902} (\bibinfo {year} {2018})}\BibitemShut {NoStop}%
\end{thebibliography}

%merlin.mbs apsrev4-1.bst 2010-07-25 4.21a (PWD, AO, DPC) hacked
%Control: key (0)
%Control: author (0) dotless jnrlst
%Control: editor formatted (1) identically to author
%Control: production of article title (0) allowed
%Control: page (1) range
%Control: year (0) verbatim
%Control: production of eprint (0) enabled
%

\onecolumngrid
\clearpage

\renewcommand{\theequation}{S\arabic{equation}}
\renewcommand{\thefigure}{S\arabic{figure}}
\renewcommand{\thesection}{S\arabic{section}}

\setcounter{equation}{0}
\setcounter{section}{0}
\setcounter{figure}{0}

\setlength{\parskip}{0.25cm}%
\setlength{\parindent}{0pt}%

\section{Neural network}
\label{sec_net}

Each network is a fully-connected architecture with $n_{\rm i}$ input nodes, $n_{\rm h}=1000$ hidden nodes, and $n_{\rm o}=2$ output nodes.  Let the indices $i\in\{0,\dots,n_{\rm i}-1\}$, $j\in\{1,\dots,n_{\rm h}\}$, and $k\in\{1,\dots,n_{\rm o}\}$ label nodes in the input, hidden, and output layers, respectively. Let $w_{\alpha \beta}$ be the weight connecting nodes $\alpha$ and $\beta$, and let $b_j$ be the bias applied to hidden-layer node $j$. Then the two output nodes take the values
\beq
S_k=n_{\rm h}^{-1} \sum_j S_j w_{jk} ,
\eeq
where
\beq
S_j= \tanh\left(\sum_i S_i w_{ij}  +b_j\right),
\eeq
and $S_i$ denotes the input-node value(s). For the time network we have $n_{\rm i}=1$ and $S_0 = t/t_0$. For the microscopic network we have $n_{\rm i}=P+1$, where $P$ is the number of patches on the disc, and $S_i$ is the number of particles in the simulation box having $i \in \{0,1,\dots,P\}$ engaged patches (divided by 1000). The mixed time-microscopic network of \f{fig_protocols} uses both $t/t_0$ and the $S_i$ as inputs. The output-node values are taken to be the changes $\Delta (\mu/\kt)$ and $\Delta (\epsilon/\kt)$, provided that $\mu/\kt$ and $\epsilon/\kt$ remain in the intervals $[-20,20]$ and $[0,20]$, respectively.

\end{document}